\documentclass[fleqn,usenatbib]{mnras}    

\usepackage{newtxtext,newtxmath}
\usepackage{gensymb}
\usepackage{chngpage}
\usepackage{amsmath}
\usepackage{longtable}
\usepackage{appendix}

\usepackage[T1]{fontenc}
\usepackage{ae,aecompl}

\usepackage{newtxtext,newtxmath}
\usepackage{gensymb}
\usepackage{amsmath}

\usepackage[T1]{fontenc}
\usepackage{ae,aecompl}

\usepackage{graphicx}	
\usepackage{amsmath}	

\usepackage{amssymb}	

\title{The Effect of Pulsar Geometry on the Observed Gamma-ray Spectrum of Millisecond Pulsars}

\author[S. J. Lloyd et al.]{
Sheridan J. Lloyd,$^{1}$\thanks{E-mail:sheridan.j.lloyd@durham.ac.uk (SJL)}
Paula M. Chadwick$^{1}$
and Anthony M. Brown$^{1}$
\\
$^{1}$ Centre for Advanced Instrumentation, Dept. of Physics, University of Durham, South Road, Durham, DH1 3LE, UK\\
\\
}
\date{Accepted 16th April 2024}

\pubyear{2024}

\begin{document}
\label{firstpage}
\pagerange{\pageref{firstpage}--\pageref{lastpage}}
\maketitle

\begin{abstract}

We analyse 13 yrs of \textit{Fermi}-LAT \textsc{PASS 8} events from 127 gamma-ray emitting millisecond pulsars (MSPs) in the energy range 0.1$-$100 GeV and significantly detect 118 MSPs. 
We fit the stacked emission with a log parabola (LP) spectral model which we show is preferred to two previously published models. 
We consider the influence of pulsar properties and observer geometric effects on spectral features by defining energy flux colours
for both the individual MSPs, and our stacked model as a baseline. There is no 
correlation of colours with pulsar luminosity, $\dot{E}$, surface magnetic field or magnetic impact angle. We also find that pulsar geometry has little effect on the observed gamma-ray spectrum which is in tension with previous modelling of gamma-ray emission with respect to pulsar geometry. Our LP MSP model is applicable to problems where an ensemble of gamma-ray MSPs is considered, such as that of the Galactic centre excess or in the case of emission from globular clusters.

\end{abstract}

\begin{keywords}
astroparticle physics: general -- gamma-rays: general -- pulsars 
\end{keywords}



\section{Introduction}

In 2009 the \textit{Fermi}-LAT established 8  millisecond pulsars (MSPs) as pulsed gamma-ray emitters using just 8 months of observations combined with radio timing data \cite{RN82}.
The initial discoveries of \cite{RN82} showed that the pulse peaks 
of radio and gamma-ray emission in MSPs need not be co-incident, favouring different emission regions for radio and gamma-rays and supporting the prevailing slot-gap and outer gap (OG) models of gamma-ray emission (\cite{Muslimov_2004} and \cite{RN731} respectively). 

The most recent fully published catalogue of gamma-ray pulsars is "The Second \textit{Fermi} Large Area Telescope Catalog of Gamma-Ray Pulsars" or 2PC, compiled by the \textit{Fermi}-LAT collaboration \cite{RN244}\footnote{A  successor catalogue to the 2PC, the 3PC was published recently and does not impact this work \cite{RN823}.}. The 2PC uses 
\textit{Fermi}-LAT observations to identify 40 MSPs above 0.1 GeV.



These MSPs have short periods, $P<$ 6 ms and spin-down slowly compared to young pulsars having period derivatives, $\dot{P}$, of 10\textsuperscript{-20} ss\textsuperscript{-1} and spin-down power, $\dot{E}$, of 10\textsuperscript{33}$-$10\textsuperscript{37} erg s\textsuperscript{-1} as compared to 10\textsuperscript{33}$-$10\textsuperscript{38} erg s\textsuperscript{-1} for young pulsars. Their observed photon fluxes are (0.2$-$9.2)$\times$10\textsuperscript{-8} cm\textsuperscript{-2} s\textsuperscript{-1} and energy fluxes (0.3$-$11)$\times$10\textsuperscript{-11} erg cm\textsuperscript{-2} s\textsuperscript{-1}. The phase-folded light curves of the MSPs most commonly exhibit 1 or 2 pulse peaks (13 and 24 MSPs respectively), although 3 MSPs exhibit 3 pulse peaks. 

The gamma-ray luminosities $L_{\gamma}$ are 10\textsuperscript{32}$-$10\textsuperscript{34} erg s\textsuperscript{-1} which indicates a typical efficiency of conversion of spin-down power to gamma-ray luminosity of $\eta<$ 1, with $\eta=L_{\gamma}/\dot{E}$. In the 2PC, $\eta$ is typically 0.1 (or 10\%) for MSPs but can be much higher, with one lower-luminosity MSP, J0610-2100 ($L_{\gamma}=10^{33}$ erg s\textsuperscript{-1}) having $\eta=10$, although this is uncommon. 

At higher energies of 10 GeV$-$2 TeV, the 3FHL catalog of  \cite{RN352} identifies 15 MSPs at Galactic latitude $|b|$ $\geq$ 10\textdegree 
~with 13 previously identified as pulsed gamma-ray emitters at lower energies using the LAT and the remaining 2 radio pulsars previously undetected in gamma-rays.

In a more recent study than the 2PC, an analysis of 19 MSPs with 7 years of \textsc{pass 8} \textit{Fermi-LAT} event data in the range 0.1$-$300 GeV shows that the empirical MSP death-line (or $\dot{E}$ below which gamma-ray emission cannot be detected) is 8$\times$10\textsuperscript{32} erg s\textsuperscript{-1} \cite{RN449}. This analysis also demonstrates that $L_{\gamma}$ is uncorrelated with $\dot{E}$, showing a nearly 2 orders of magnitude of scatter for MSPs with an $\dot{E}$ of 10\textsuperscript{34} erg s\textsuperscript{-1}. For $\dot{E}>$  5$\times$10\textsuperscript{34} erg s\textsuperscript{-1}, $L_{\gamma}$ is directly proportional to $\sqrt{\dot{E}}$.  Similarly the wide variation of  the relationship between $\eta$, $L_{\gamma}$ and $\dot{E}$ makes it difficult to deduce gamma-ray luminosities \textit{a priori} from MSP timing information alone.

In contrast, the geometry of individual MSPs is increasingly well understood, through the fitting of models to gamma-ray and radio light curves, with the angles for both the line of sight and magnetic axis, with respect to the pulsar spin axis determined, for all 40 MSPs in the 2PC \cite{RN28} and more recently for 10 MSPs with high signal-to-noise observations \cite{RN635}. 

We note that there is recent modelling and interest in the effects of pulsar inclination (among other variables) on the observed gamma-ray spectra of MSPs ranging from a general one \cite{RN376,RN451}, to a spectral softening below 400 MeV, to being a lesser influence on the high-energy spectrum \cite{RN452,RN716}. Thus, the effect of pulsar magnetosphere geometry and magnetic inclination angle on the observed gamma-ray spectra is an open question. Therefore, inclination effects could well affect the the observed gamma-ray spectrum, although to our knowledge this has not been confirmed by observation.

A demonstrable link between pulsar inclination and observed spectra could serve as a constraint on spectral emission models where the pulsar geometry is known, whilst conversely the geometry of the pulsar system may be constrained from the observed gamma-ray spectrum. We are therefore motivated in this paper to determine for the first time if there is a link between pulsar geometry and \textit{observed} gamma-ray spectral features.

In this work, we seek to construct the best general spectral model for 118 MSPs and use it to examine the effect of geometry on gamma-ray spectra, whilst excluding any systematic effect of other pulsar properties (surface magnetic field, luminosity , $\dot{E}$). We analyse a larger sample of 127 MSPs, rather than the 39 MSPs of the 2PC \cite{RN244}, as used by the previous stacked MSP analyses of~ \cite{RN183} and  \cite{RN197}. Furthermore, we note that the MSP spectral model we produce is very useful in its own right, and could be applied to any problem where gamma-ray emission is presumed to originate from an ensemble of MSPs plus another component such as emission from dark matter, as relevant to globular clusters or the Galactic centre excess .

This paper is structured as follows. In Section \ref{sec:GCSelection} we describe the selection criteria of MSPs for analysis. In Section \ref{sec:CH5Analysis} we describe our analysis method for the detection of MSPs, and in  Section \ref{sec:AnalysisResults} we present our results. In Section \ref{sec:StackedModel} we derive a spectral model for the stacked emission of detected MSPs. In Appendix \ref{sec:DiscussionModels} we compare our model to those of \cite{RN183} and  \cite{RN197} and determine whether there is any relationship between colours constructed from low and medium energy MSP fluxes, and pulsar $\dot{E}$, luminosity and surface magnetic field. We also consider the spectral predictions of synchro-curvature emission models incorporating pulsar magnetic inclination and determine if they are suitable for comparison with the observed gamma-ray spectra. Finally in Section \ref{sec:Conclusion} we summarise our findings and make suggestions for future work.
   
\section{Millisecond Pulsar Selection}
\label{sec:GCSelection}

We select all 127 MSPs (defined as any pulsar with period  $\leq$ 30 ms) from the Public List of LAT Detected Gamma-Ray Pulsars 
\footnote{https://confluence.slac.stanford.edu/display/GLAMCOG/ \\ Public+List+of+LAT-Detected+Gamma-Ray+Pulsars, accessed on 24\textsuperscript{th} August 2021}. Their co-ordinates, periods and $\dot{E}$ values are shown in Table~\ref{tab:MSP_SELECTION_LIST}. 

\section{Analysis Method}
\label{sec:CH5Analysis}
\subsection{Initial Photon Event Data Selection}
\label{sec:CH5PhotonEventDataSelection} 
The data in this analysis were collected by \textit{Fermi}-LAT between 4th Aug 2008 (15:43 h) to 26th August 2021 (00:52 h), (Mission Elapsed Time (MET)  2395574147[s] to 651631981[s]). 
We select all \textsc{pass} 8 events which are \textit{source} class  photons (evclass = 128), and PSF3 events with the best quartile direction reconstruction, (evtype = 32), spanning the energy range 0.1$-$100 GeV. Throughout our analysis, the \textit{Fermipy} software package\footnote{\textit{Fermipy} change log version 1.0.1 (\cite{2017arXiv170709551W})} with version \textsc{2.0.8} of the \textit{Fermitools} is used, in conjunction with the \textsc{p8r3\_source\_v3} instrument response functions. We apply the standard \textsc{pass}8 cuts to the data, including a zenith angle 90\textdegree{} cut to exclude photons from the Earth limb, and good-time-interval cuts of DATA\_QUAL > 0 and LAT\_CONFIG = 1. The energy binning used is 4 bins per decade in energy and spatial binning is 0.1\textdegree{} per image pixel. 

\subsubsection{Spectral Models}

The differential flux, \textit{dN/dE}, (photon flux per energy bin) of an individual MSP, is described by a spectral model which is either an exponential cut-off power law 2  (Eqn.~\ref{PLexp}, PLSuperExpCutoff2 ), a log parabola (Eqn.~\ref{LPeqn}, LP), or a power law ( Eqn.~\ref{PLeqn}, PL) \footnote{As described in the FSSC source model \protect\cite{RN181}}.

\begin{equation}{\label{PLexp}}
\frac{dN}{dE}=N\textsubscript{0} \Big(\frac{E}{E_0}\Big)^{\gamma_1} \exp\Big((-aE)^{\gamma_2}\Big)
\end{equation}
                
where \textit{normalisation} (also known as \textit{prefactor}) = $N$\textsubscript{0}, \textit{index1} = $\gamma_1$,  $E$\textsubscript{0} is the scale, $a$ is the exponential factor and \textit{index2 } = $\gamma_2$ (which controls the sharpness of the exponential cut-off).

\begin{equation}{\label{LPeqn}}
\frac{dN}{dE}=N\textsubscript{0}\Big(\frac{E}{E\textsubscript{b}}\Big)^{-(\alpha+\beta \log ({E}/{E\textsubscript{b}}))}\
\end{equation}
where \textit{norm} = $N$\textsubscript{0}, 
and  $E$\textsubscript{b} is a \textit{scale} parameter.

\begin{equation}{\label{PLeqn}}
\frac{dN}{dE}=N\textsubscript{0}\Big(\frac{E}{E\textsubscript{0}}\Big)^\gamma\
\end{equation}
where \textit{prefactor} = $N_{0}$, \textit{index} = $\gamma$ and \textit{scale} = $E_{0}$.

\subsection{MSP Likelihood Analysis}

We perform a full likelihood analysis  on all 127 MSP targets (Table~\ref{tab:MSP_SELECTION_LIST}), in the energy range 0.1$-$100 GeV, using a 25\textdegree{} Radius of Interest (ROI) centred on the nominal MSP co-ordinates and a 40\textdegree{} Source Radius of Interest (SROI) for each MSP target. The model we use in our likelihood analysis consists of a point source population seeded from the \textit{Fermi}-LAT's 10 yr fourth point source catalog, data release 2, (4FGL-DR2, gll\_psc\_v27.fit ), extended gamma-ray sources and diffuse gamma-ray emission. The diffuse emission detected by the \textit{Fermi}-LAT consists of two components: the Galactic diffuse flux, and the isotropic diffuse flux. The Galactic component is modelled with \textit{Fermi}-LAT's gll\_iem\_v07.fit spatial map with the normalisation free to vary. The isotropic diffuse emission is defined by \textit{Fermi's}  iso\_\textsc{P8R3}\_\textsc{SOURCE}\_\textsc{V3}\_PSF3\_v1.txt tabulated spectral data. The normalisation of the isotropic emission is left free to vary.

In addition, all known sources (including MSPs) take  initial spectral parameters and position from the 4FGL.

Initially, the \textit{setup} and \textit{optimize} methods are run to create count and photon exposure maps and to compute the test statistic (TS) values of all 4FGL sources in the model. The TS value is defined as Eqn.~\ref{MATTOX_TS} where $L_0$ is the maximum likelihood for a model without a source observations (i.e. the null hypothesis) and $L_1$ is the maximum likelihood for an additional source observation at a given location.

\begin{equation}{\label{MATTOX_TS}}
TS \equiv -2(\ln L_0-\ln L_1)
\end{equation}

The \textit{fit} method is then run: \textit{fit} is a likelihood optimisation method which executes a fit of all parameters that are currently free in the the model and updates the TS and predicted count ($Npred$) values of all sources. From this fit, all point sources with a TS $<4$, or with a predicted number of photons $<4$ are removed from the model.

The normalisation of all sources within 10\textdegree{} of the MSP are then freed using  the \textit{free\_source} method to allow for the Point Spread Function (PSF) of PSF3 events down to 100 MeV (95~\% containment at 8\textdegree)
. The source nearest to the catalogue position of the MSP  has \textit{prefactor} and \textit{index} spectral parameters (Eqn.~\ref{PLeqn}) freed for power law sources and  \textit{prefactor}, \textit{index1} and \textit{expfactor} freed for PLSuperExpCutoff2 sources (Eqn.~\ref{PLexp}). 

The shape and normalisation parameters of all sources with a TS > 25 are then individually fitted using the \textit{optimize} method. Then the \textit{fit} method is run twice more with an intervening \textit{find\_sources} step.  \textit{Find\_sources} is a peak detection algorithm which analyses  the test statistic (TS) map to find new sources over and above those defined in the 4FGL 
model by placing a test point source, defined as a power law (PL) with spectral index 2.0, at each pixel on the TS map and recomputing likelihood. Finally, the \textit{sed} method generates a spectral energy distribution, with energy dispersion disabled for all MSPs which are known 4FGL sources, and a 2$\sigma$ confidence limit on the determination of instrument upper limits.

\section{Analysis Results}
\label{sec:AnalysisResults}
We detect 118 of the 127 catalogue MSPs in the energy range 0.1$-$100 GeV at a significance of 5$\sigma$ (test statistic, TS$=$25) or greater. 
The 9 MSPs, J0154+1833, J0636+5129, J1327-0755, J1946+3417, J1455-3330, J1909-3744, J1832-0836, J2205+6012 and J2317+1439 are undetected through their integrated emission in our study, but 6 of these are detectable in other work by phase folded pulsed emission using a radio ephemeris \cite{RN459,smith2017gammaray}. We list the TS, offset from catalogue co-ordinates, and energy and photon fluxes in Table~\ref{tab:MSP_PL_FLUXES}.

We also list the best fit spectral models for MSPs with a PLSuperExpCutoff2, LP or PL spectral model, in Tables~\ref{tab:MSP_PLSUPEREXP_2_MODELS}, \ref{tab:MSP_LP_MODELS} and \ref{tab:MSP_PL_MODELS}, respectively. For these spectral models we provide a selection of representative spectral energy distributions in Figs.~\ref{fig:PLSUP_EXP_2_SEDS} and ~\ref{fig:PL_LP_SED}.


We determine the average differential energy flux per energy bin (for flux points of $\geq$ 2$\sigma$ significance) and the standard error of the mean for the detected MSPs (Table~\ref{tab:MSP_SEM_COUNT}). The energy bin centres and lower and upper bin energy (extent) arise from a binning of 4 bins per decade of energy as specified in the analysis. Note that the  error (energy dispersion) of the bin extent is $<$ 10~\% between 1$-$100 GeV increasing to 20~\% at 0.1 GeV.

The majority of MSPs display emission across the whole energy range between 0.2$-$10 GeV, while  emission between 10$-$18 GeV is seen in only 41~\% of MSPs, and 15~\% of MSPs above 18 GeV.  

\begin{table*}
 \centering
	
	\begin{tabular}{lc c c c c } 
		\hline
                Bin Centre	&  Lower Bin Energy &Upper Bin Energy &  MSP Count and Percentage of Sample   & Mean Energy Flux 		\\
                GeV         &  GeV           & GeV      & with Significant Flux in Bin            & 10\textsuperscript{-6} MeV cm\textsuperscript{-2} s\textsuperscript{-1}  \\
		\hline
		

0.13	&	0.10	&	0.18	&	28	(	23.7	)	&	2.23	$\pm$	0.34	\\
0.24	&	0.18	&	0.32	&	85	(	72.0	)	&	2.30	$\pm$	0.20	\\
0.42	&	0.32	&	0.56	&	106	(	89.8	)	&	2.49	$\pm$	0.24	\\
0.75	&	0.56	&	1.00	&	112	(	94.9	)	&	2.77	$\pm$	0.29	\\
1.33	&	1.00	&	1.78	&	114	(	96.6	)	&	2.97	$\pm$	0.32	\\
2.37	&	1.78	&	3.16	&	116	(	98.3	)	&	2.71	$\pm$	0.30	\\
4.22	&	3.16	&	5.62	&	112	(	94.9	)	&	2.21	$\pm$	0.28	\\
7.50	&	5.62	&	10.00	&	92	(	78.0	)	&	1.37	$\pm$	0.22	\\
13.34	&	10.00	&	17.78	&	48	(	40.7	)	&	0.87	$\pm$	0.17	\\
23.71	&	17.78	&	31.62	&	18	(	15.3	)	&	0.73	$\pm$	0.17	\\
42.17	&	31.62	&	56.23	&	7	(	5.9	)	&	0.56	$\pm$	0.11	\\
74.99	&	56.23	&	100.00	&	1	(	0.8	)	&	0.09	$\pm$	0.00	\\

		\hline
	\end{tabular}
    	\caption{The count and percentage of significant ($>2\sigma$) flux points for detected MSPs along with mean flux and associated standard error of the mean. 
    	}
        \label{tab:MSP_SEM_COUNT}
\end{table*}

\section{GEOMETRIC EFFECTS ON THE OBSERVED GAMMA-RAY SPECTRA OF MILLISECOND PULSARS}
\label{sec:GEOMETRIC}

The geometry of a pulsar system is described by (1) the angle between the observer line-of-sight (LOS) and the pulsar spin axis($\zeta$), (2) the angle between magnetic axis and the pulsar spin axis, also known as the magnetic obliquity ($\alpha$) and (3) the orbital inclination angle between the pulsar orbital plane and the observer LOS ($i$). An impact parameter ($\beta$) describing the closest approach of the magnetic axis and the LOS is defined as $\beta=\zeta-\alpha$ as in \cite{RN28}. 

In \cite{RN458}, $i$ is derived for Galactic field MSPs, with helium white dwarfs as binary companions, from the relationship of pulsar and white dwarf mass, and the orbital period. Over the life of the system, the values of  $i$ and $\zeta$ with respect to the pulsar rotation axis are shown to align, with the detection of gamma-ray emission in Galactic MSPs appearing to be somewhat favoured by smaller values of $\zeta$. 

In \cite{RN376} the effect of parameters and assumptions in the OG model are examined. The magnetic inclination angle of a pulsar is shown to affect the magnetosphere geometry, position of the OG and radius of curvature (\textit{r\textsubscript{c}}). \textit{r\textsubscript{c}} is identified as the most important parameter having an order of magnitude impact on the parallel electric field, (responsible for lepton acceleration and consequent curvature radiation), which is noted to affect the observed gamma-ray spectrum qualitatively.  

In their follow-up paper, \cite{RN451} model exemplar pulsar gamma-ray spectra obtained by integrating the energy loss of a single particle with distance travelled, convolved with an effective observed particle distribution which incorporates geometry and beaming effects. They show that photons from the initial part of the particle trajectory exhibit softer low energy spectra below 200 MeV, with an index of 0.68 vs an index of 1.14 for emission from the whole trajectory. They note that such spectra could be observed for pulsars with a favourable viewing geometry of the outer gap.  

 The effect of inclination on the observed gamma-ray spectrum is also considered by \cite{RN522} who model the combined low-energy synchro-curvature radiation from the polar cap and higher-energy pulsar synchrotron radiation (HES) from the equatorial current sheet (ECS) using a particle-in-cell (PIC) simulation. They produce phase-averaged observed pulsar energy spectra between 400 keV $-$ 4 GeV for varying $\alpha$ and $\zeta$. These spectra show a trend of decreasing energy flux and a softening of the spectrum between 40$-$400 MeV, with the disappearance of the HES spectral peak at $\sim$ 200 MeV as $\beta$ decreases from 90 to 0\textdegree~ for $\zeta$ = 90\textdegree.~ In contrast, varying $\alpha$ from 0 to 90\textdegree~ for $\zeta$ angles of 45 and 60\textdegree~ generally results in few spectral shape changes.

In \cite{RN347}, a minimalist model determines lepton particle trajectories and velocities depending only on local $B$ field using the vacuum rotator dipole model (VRDM) of pulsars as first described by \cite{Deutsch}. The gamma-ray spectrum arising from the resulting curvature radiation is determined and used to produce sky-maps and light curves for varying pulsar phase, angle $\zeta$ and magnetic inclination angle ($\chi$ in that paper rather than $\alpha$), with respect to the pulsar rotation axis. This modelling is able to reproduce spectral cut-offs at a few GeV and single or double peaked light curves broadly consistent with \textit{Fermi}-LAT observations of pulsars. It is also shown that there is a slight hardening of the spectrum between 1$-$10 GeV, when the annulus of the emitting region increases in extent, as it moves closer to the surface of the star, from 0.5$-$1 of the light cylinder radius ($R_{LC}$) to 0.1$-$1 $R_{LC}$. Beyond the light cylinder there is no change in emission spectrum with increasing extent (1$-$5 $R_{LC}$).  Overall the spectral shape is noted to be insensitive to geometry.

In \cite{RN452}, a synchrocurvature spectral emission model is able to reproduce well the observed spectral energy distribution of 18 out of 32 non-MSPs\footnote{Although their selection criteria for a non-MSP is a pulsar with period $>$ 10 ms, their 18 pulsars all have period $>$ 30 ms and thus also meet our definition of a non-MSP.} considered, from X-ray to gamma-ray energies. The free parameters of the model are properties intrinsic to the pulsar, namely, the parallel electric field and the magnetic field gradient, and properties related to the viewing geometry of the observer: a length scale for particle emission relative to the radius of the light cylinder and a normalisation of the accelerated particle distribution with respect to distance ($x_0/R_{lc},N_0$ respectively).  This model can reproduce the gamma-ray spectrum of the 4 MSPs considered but not the X-ray spectrum, whilst 3 other MSPs can be well fitted simultaneously in X-rays and gamma-rays. Extending the model so that there are two emission regions (each with own values of ($x_0/R_{lc},N_0$, but same intrinsic properties) and visible to a varying extent depending on observer line-of-sight, is shown to fit both the X-ray and gamma-ray observations. This suggests that, whilst broad-band emission from X-rays to gamma-rays can be sensitive to geometric effects and the visibility of different emission regions, gamma-ray emission is less so. The authors also note that the precise location and extent of the accelerating region has previously being shown not to dominate the predicted high-energy spectral shape.

Finally, \cite{RN716} model the gamma-ray emission arising from synchrotron and curvature radiation in high-altitude slot gaps at the separatrix between open and closed field lines in the pulsar magnetosphere for pulsars of period 2 ms. They use a VRDM with a mono-energetic lepton distribution to produce phase-integrated and LOS averaged photon count spectra between 1$-$30 GeV, with spectral shape and energy limits that vary with magnetic obliquity, $\chi$, and LOS, $\zeta$, whilst retaining a spectral peak at around 4 GeV. Specifically, increasing the magnetic obliquity (for phase and line of sight averaged spectra) from 30 to 90 \textdegree ~increases spectral energy limits from 7 to 30 GeV.  However, they conclude that, while the pulsar radio spectra significantly depend on the magnetic obliquity, the high-energy part of the spectrum is much less sensitive to geometry. 

The models of \cite{RN522,RN347,RN452, RN716} are broadly similar in that they (1) derive the curvature radiation spectrum using a Bessel function of order 5/3, (2) are force-free allowing leptons to travel along magnetic field lines, because the screening of the pulsar electric field by plasma is excluded and (3) balance lepton acceleration and braking due to radiative friction. \cite{RN347} and \cite{RN716} employ a mono-energetic particle distribution whereas \cite{RN452} spatially vary the number density of particles.


For the models above, we summarise the effect of geometry on the gamma-ray spectrum in Table~\ref{tab:INCLINATION_EFFECTS}, varying from no effect to changes in energy limits and spectral shape with varying $\alpha$. 
However, the models which present explicit spectra for specific geometries are difficult to compare directly with gamma-ray observations of MSPs, as they either do not provide a full co-varying range of $\zeta$ and $\alpha$ (\cite{RN716}), or are designed to test a model concept rather than to be fitted directly to observations (\cite{RN522}), and moreover they do not reproduce gamma-ray spectral peaks of a few GeV as seen in MSPs.

\begin{table}
 \centering
	
        \begin{tabular}{p{0.3\linewidth} | p{0.5\linewidth}}
		\hline
                Model & Inclination Effect    \\
                Reference      & on MSP gamma-ray Spectrum \\
		\hline

\cite{RN347}& Overall, the spectral shape is insensitive to geometry.     \\
\hline
\cite{RN452}& Gamma-ray spectrum insensitive to geometry but two visible emission regions needed to account for both X-ray and gamma-ray emission.      \\
\hline
\cite{RN376}&Qualitatively, magnetic inclination affects spectrum via influence on radius of curvature and the consequent parallel electric field.     \\
\hline
\cite{RN451}&Favourable viewing geometry sampling initial particle trajectories produces softer low-energy spectra.      \\
\hline
\cite{RN522}&A proof of concept model showing spectral softening below 400 MeV for $\zeta$=90\textdegree~as $\alpha$ increases, in contrast with few spectral shape changes for $\zeta$ = 45 and 60\textdegree.      \\
\hline

\cite{RN716}&Increasing magnetic inclination increases upper energy limit of $\zeta$ averaged photon spectra from 7 to 30 GeV but high-energy spectra noted to be much less sensitive to geometry than radio bands.      \\

		\hline
	\end{tabular}
    	\caption{Summary of inclination effects on gamma-ray spectra as indicated by current models, varying from no effect, to changes in model spectra as magnetic obliquity, $\alpha$,  varies}
        \label{tab:INCLINATION_EFFECTS}
\end{table}

As no model spectra are suitable for comparison with observations, we instead consider the observed MSP gamma-ray spectrum to investigate if geometry has any systematic effect.

Specifically, we would like to examine the most rapidly varying part of the MSP spectrum, which is generally away from the cut-off values of the the exponential power law in the 2PC (1.2$-$5.3 GeV), where the spectrum peaks and is often almost flat. Moreover, we wish to probe energy ranges which are noted as being sensitive to emission extent and geometric effects as in Table \ref{tab:INCLINATION_EFFECTS} above.

We therefore define two colour band ratios to examine the requisite energy ranges, a low-energy ratio, (Eqn.~\ref{LE_COLOUR}, LE)  and a medium energy ratio, (Eqn.~\ref{ME_COLOUR}, ME). 

\begin{equation}{\label{LE_COLOUR}}
LE=\frac{Energy~Flux\textsubscript{~133 MeV}}{Energy~Flux\textsubscript{~237 MeV}}
\end{equation}

\begin{equation}{\label{ME_COLOUR}}
ME=\frac{Energy~Flux\textsubscript{~4217 MeV}}{Energy~Flux\textsubscript{~7499 MeV}}
\end{equation}

The precise energy centres and limits for LE and ME arise from the binning employed, but the energy bins are also chosen to examine the varying and potentially inclination-sensitive part of the MSP spectrum as noted above. If an upper limit (UL) is present at 133 MeV or 7498 MeV, then the LE and ME colours represent upper and lower limits respectively. 

 If a decreasing value of $\beta$ results in an increasingly soft spectrum below 400 MeV as in \cite{RN522} then we should expect a corresponding increase in the LE colour as $\beta$ decreases from 90 to -45 \textdegree. Conversely, if geometric effects systematically affect the spectrum between 1$-$10 GeV then this should appear as a changing ME colour as $\beta$ changes,  with the caveat that increasing emission extent from within the light cylinder could result in a decreasing ME colour. 
 
\label{sec:StackedModel}
\section{The Stacked MSP Spectrum}

Whilst we consider the effect of geometry on the spectra of individual MSPs, these spectra can vary between MSPs, and so we must derive a stacked MSP model to serve as a baseline to show if there are systematic spectral changes depending on geometry across the MSP sample. In order to derive this model for an ensemble of Galactic field MSPs, we sum the energy fluxes ($E^2\frac{dN}{dE}$) in each bin (ignoring ULs) between 0.1$-$56.2 GeV for MSPs detected in the energy range 0.1$-$100 GeV, with the upper energy range of 56.2 GeV arising solely from the binning employed. We exclude the 1 flux point above 56.2 GeV (for the single MSP detection J1903-7051) from the stacked model, as this flux is only marginally significant (TS 5.5).

\begin{figure}
	
	\includegraphics[width=\columnwidth]{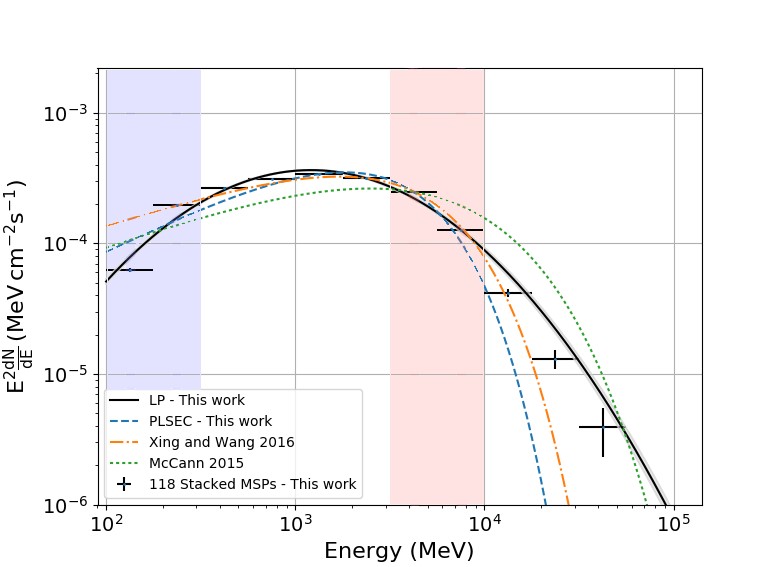}
	\centering
    \caption{The likelihood best-fit models for LP and PLSEC (both this work), McCann, and Xing and Wang for the stacked spectrum of 118 MSPs. The LP model is the best fit overall and has just visible 1$\sigma$ uncertainty indicated by grey shading. The blue and red bands (each comprising two bins), indicate the overall energy range used to form the low and medium energy colours respectively as in Eqns.~\ref{LE_COLOUR} and ~\ref{ME_COLOUR}. For clarity the uncertainty of the McCann (1$\sigma$) and the Xing and Wang (3$\sigma$) models is not shown.}
    \label{fig:replace_bands}
\end{figure}

We define two models for comparison; a super exponential cut-off power law model spectral model as Eqn.~\ref{PLexp}, "PLSEC" and a log parabola model as Eqn.~\ref{LPeqn}, "LP". 

In the PLSEC model, the parameters \textit{index1}, \textit{Normalisation} and \textit{exponential factor} are left free. The \textit{scale} and \textit{index2} parameters are frozen to 1$\times$10\textsuperscript{-3} TeV and 1 respectively. In the LP model, the \textit{Norm}, $\alpha$ and $\beta$ parameters are left free, while $E$\textsubscript{b} is frozen at 1 TeV.

We then use \textsc{gammapy} version 0.18.2\footnote{Available from https://docs.gammapy.org/0.18.2/} software to fit the flux summation of 118 MSPs with the LP and PLSEC models in the range 100 MeV$-$56.2 GeV. 

In Appendix~\ref{sec:DiscussionModels} we show that the LP model is the best model to describe the stacked spectrum of the 118 MSPs and preferred to the previously published models of \cite{RN183} and \cite{RN197}. The parameters of our best fit LP model 
are presented in Table~\ref{tab:LP_MODEL_PARAMS} and all models are presented alongside the stacked MSP spectrum in In Fig.~\ref{fig:replace_bands}.

\begin{table}
 \centering
	
	\begin{tabular}{lc c c} 
		\hline
                Parameter	& Value & Unit \\
		\hline

$\alpha$&  1.88 $\ \pm~0.01 $ & - \\
$\beta$& (3.16$\ \pm \ {0.04} \ )\ \times \ $10\textsuperscript{-1} & - \\
\textit{Normalisation}& (3.49$\ \pm \ {0.02} \ )\ \times \ $10\textsuperscript{-4}  & cm\textsuperscript{-2} s\textsuperscript{-1} TeV\textsuperscript{-1} \\
\textit{Scale}&  1$\ \times \ $10\textsuperscript{-3} & TeV  \\

		\hline
	\end{tabular}
    	\caption{The parameters of the best-fit LP model (this work) using Eqn.~\ref{LPeqn}, for the stacked differential energy flux of 118 significantly detected MSPs in the energy range 100 MeV$-$56.2 GeV. The LP model is the best fit overall compared to all other models considered.}
        \label{tab:LP_MODEL_PARAMS}
\end{table}

The LE and ME colours for this LP model are 0.50$\pm$0.02 and 1.75$\pm$0.05 respectively. These serve as a baseline model to determine spectral trends for individual MSPs in our selection. We show the energy range used to form the LE and ME colour ratios on Fig.~\ref{fig:replace_bands}.

\onecolumn
\begin{figure}
\centering

\includegraphics[width=0.49\textwidth]{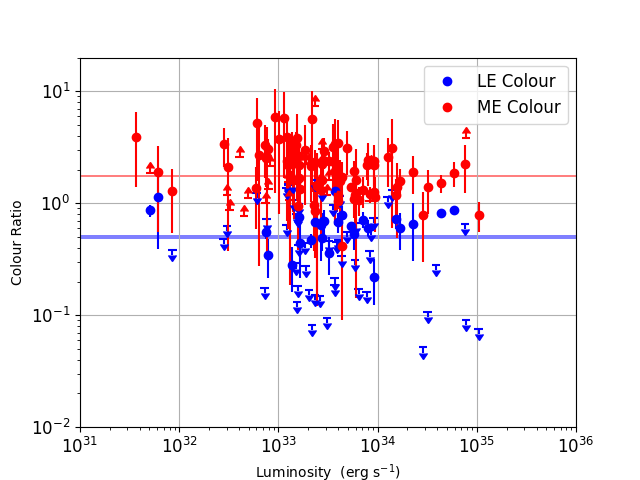}
\includegraphics[width=0.49\textwidth]{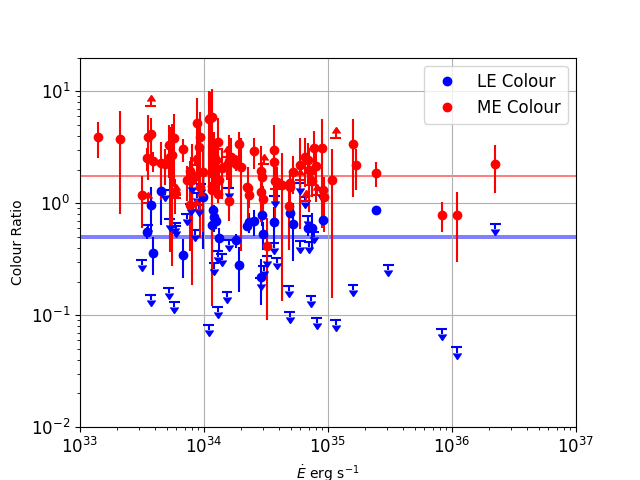}
\includegraphics[width=0.49\textwidth]{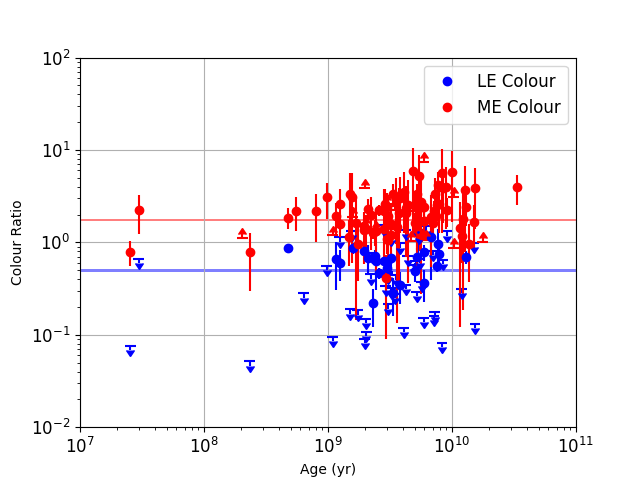} 
\includegraphics[width=0.49\textwidth]{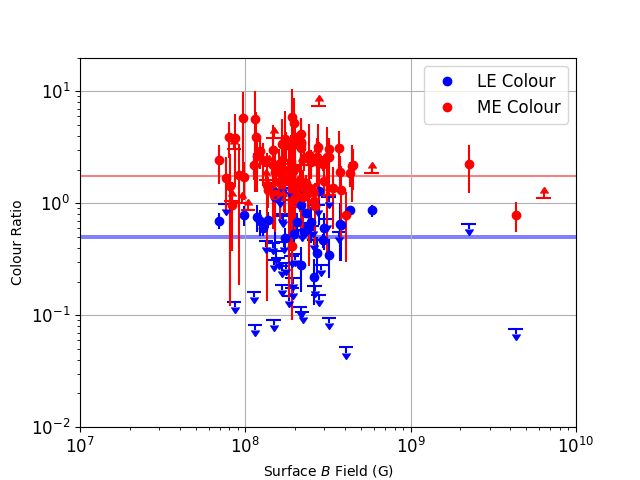} 

    \caption{
     The LE and ME colours for the 118 MSPs plotted against pulsar luminosity, $\dot{E}$, characteristic age and surface magnetic ($B$) field. There is no apparent correlation with LE and ME colours distributed evenly above and below the LP colour values (red and blue bands). Colour ratios are shown either as points with errors (where flux points exist in all bins for the colour), or as upper or lower limits where there are a mixture of upper limit and flux point observations in the energy bands used to form the colour ratios.
    }
\label{fig:colour_msp_key_props}   
\end{figure}

\begin{figure}
\includegraphics[width=0.5\textwidth]{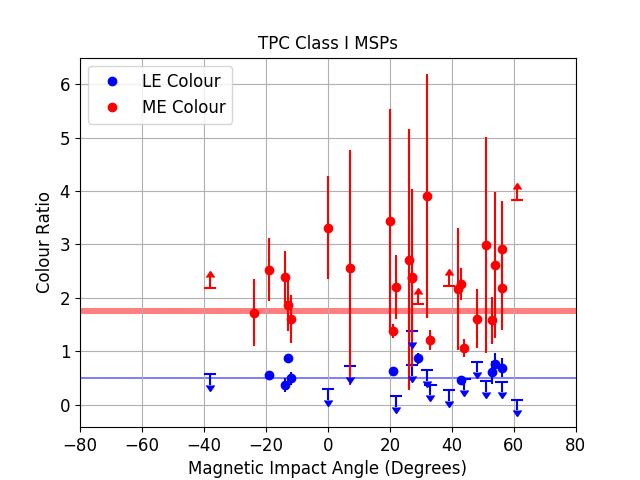}
\includegraphics[width=0.5\textwidth]{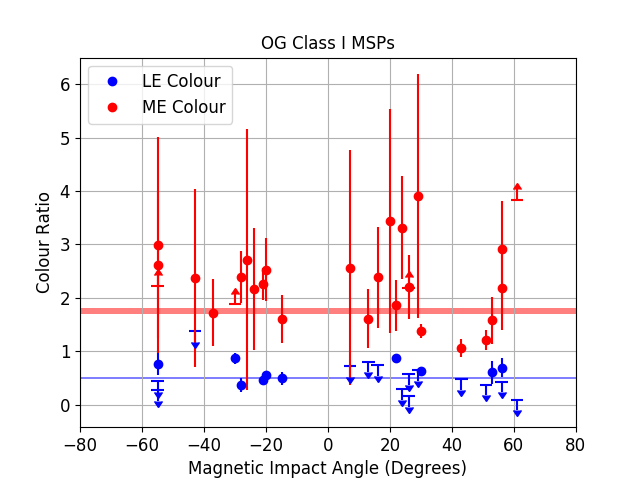}
\\
    \caption{The LE and ME colours for Class I MSPs vs a magnetic impact angle ($\beta$) from inclination angles in \protect\cite{RN28} for a two-pole caustic model (TPC), and an outer-gap (OG) model. There is no apparent correlation of colour and impact angle. The LP colour values are shown as red and blue bands. Colour ratios are shown either as points with errors, or as upper or lower limits.    }
\label{fig:RN28_CLASS_I_MODELS}   
\end{figure}

\twocolumn
In considering the effect of inclination on spectra, we firstly need to exclude any correlation of LE and ME colours with intrinsic pulsar properties such as $\dot{E}$, luminosity or characteristic age ($P/2\dot{P}$). A plot of MSP spectral colours against luminosity, $\dot{E}$, age and surface magnetic field indicates that there is no detectable correlation (Fig.~\ref{fig:colour_msp_key_props}, Pearson correlation coefficient for ME colours of -0.23, -0.13, -0.30 and -0.16 respectively)

\begin{figure}
\includegraphics[width=0.5\textwidth]{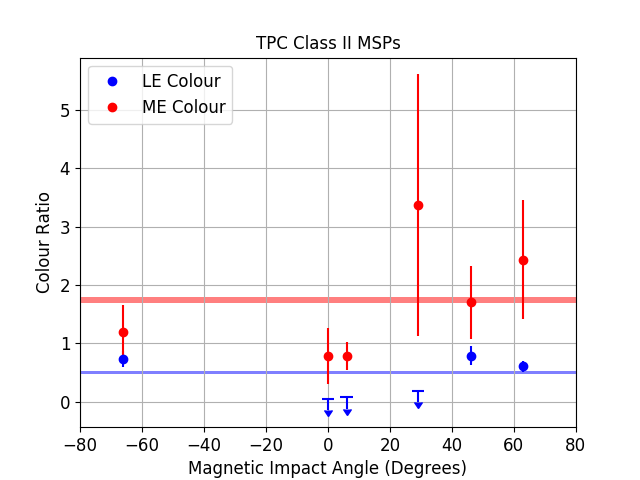}
\includegraphics[width=0.5\textwidth]{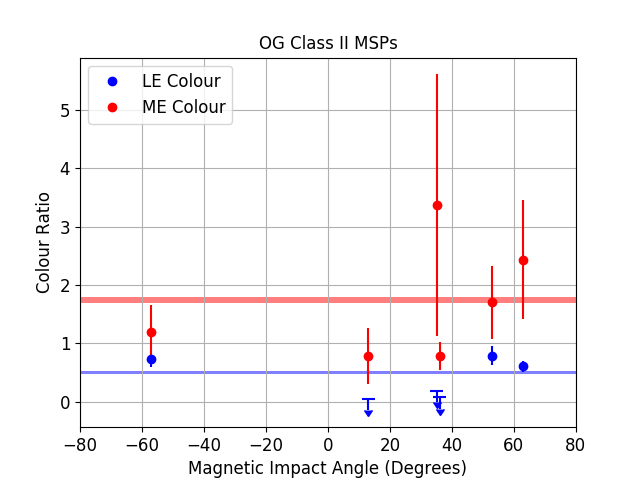}
\includegraphics[width=0.5\textwidth]{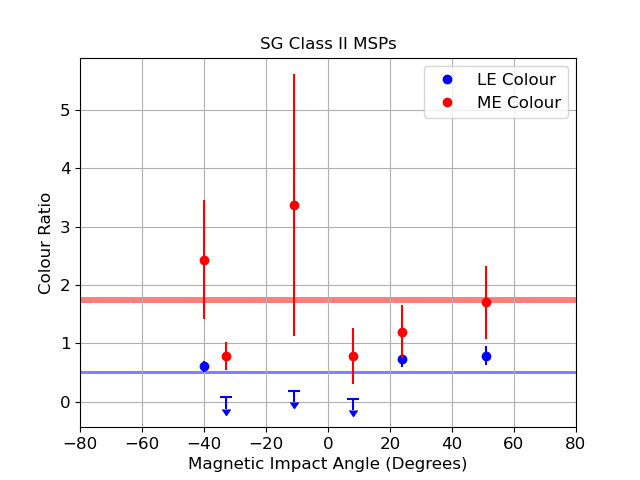}
\\
    \caption{  The LE and ME colours for Class II MSPs vs a magnetic impact angle ($\beta$) determined from inclination angles in \protect\cite{RN28} for a two-pole caustic  (TPC), an outer-gap (OG) and a slot-gap (SG) model. There is no apparent correlation of colour and impact angle. The LP colour values are shown as red and blue bands. Colour ratios are shown either as points with errors, or as upper or lower limits.}
    
\label{fig:RN28_CLASS_II_MODELS}   
\end{figure}

\begin{figure}
\includegraphics[width=0.5\textwidth]{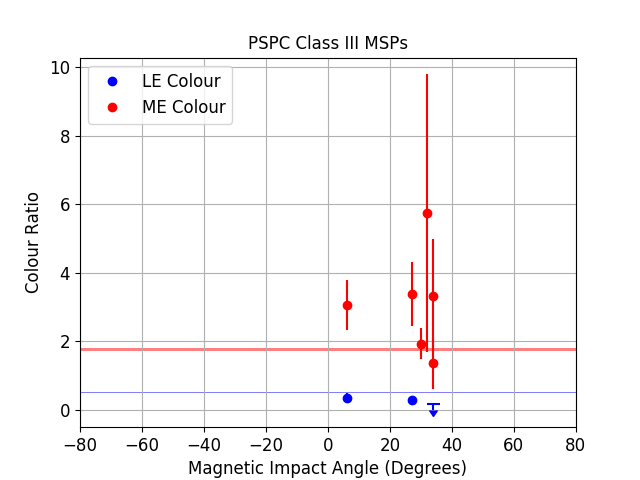}
\\
    \caption{
    The LE and ME colours for Class III MSPs vs a magnetic impact angle ($\beta$) determined from inclination angles in \protect\cite{RN28}, for a pair-starved polar cap (PSPC) model.  There is no apparent correlation of colour and impact angle. The LP colour values are shown as red and blue bands. 
    }
\label{fig:RN28_CLASS_III_MODELS}   
\end{figure}

We next consider whether the inclination (geometric) properties of the MSP, such as the LOS, $\zeta$, and the magnetic inclination angle, $\alpha$, with respect to the pulsar rotation axis might affect spectral features. \cite{RN28} use a VRDM to derive the best fit $\zeta$ and $\alpha$ values, for the MSPs of the 2PC. They divide the MSPs into 3 classes, Class I, II and III, defined as the MSP's gamma-ray pulse trailing, aligned or leading the MSP's radio pulse, respectively. They then derive inclination angles from a two-pole caustic (TPC) and outer-gap (OG) model for Class I MSPs, a TPC, OG and slot-gap model for Class II MSPs and a pair-starved polar cap model for Class III. We use these magnetic inclination angles to determine the magnetic impact angle, $\beta$, defined as $\beta=\zeta-\alpha$ for the three classes of MSPs (Tables  ~\ref{tab:RN28_TABLE_6_AND_7},~\ref{tab:RN28_TABLE_8_AND_9_AND_10} and ~\ref{tab:RN28_TABLE_11}). We then plot $\beta$ for individual MSP colours in Fig.~\ref{fig:RN28_CLASS_I_MODELS},~\ref{fig:RN28_CLASS_II_MODELS}, ~\ref{fig:RN28_CLASS_III_MODELS}. We observe that there is no dependence of ME and LE colours on $\beta$, implying that LE and ME gamma-ray spectral features of MSPs are not correlated with magnetic inclination and line-of-sight effects.

\begin{table*}
 \centering
	
	\begin{tabular}{lc c c  c c c c } 
		\hline
&		&	TPC	&		&		&	OG	&		\\

MSP	&	$\alpha$ (\textdegree)&	$\zeta$	(\textdegree)&	$\beta$	(\textdegree)&	$\alpha$ (\textdegree)&	$\zeta$	(\textdegree)&	$\beta$	(\textdegree)\\

        \hline
        
J0023+0923	&	38	&	65	&	27	&	67	&	24	&	-43	\\
J0030+0451	&	74	&	55	&	-19	&	88	&	68	&	-20	\\
J0101-6422	&	30	&	84	&	54	&	90	&	35	&	-55	\\
J0102+4839	&	43	&	70	&	27	&	60	&	76	&	16	\\
J0218+4232	&	25	&	12	&	-13	&	45	&	67	&	22	\\
J0437-4715	&	35	&	64	&	29	&	76	&	46	&	-30	\\
J0610-2100	&	87	&	49	&	-38	&	63	&	89	&	26	\\
J0613-0200	&	55	&	43	&	-12	&	60	&	45	&	-15	\\
J0614-3329	&	63	&	84	&	21	&	58	&	88	&	30	\\
J0751+1807	&	21	&	69	&	48	&	59	&	72	&	13	\\
J1024-0719	&	66	&	73	&	7	&	66	&	73	&	7	\\
J1124-3653	&	13	&	69	&	56	&	17	&	73	&	56	\\
J1125-5825	&	29	&	71	&	42	&	84	&	60	&	-24	\\
J1231-1411	&	26	&	69	&	43	&	88	&	67	&	-21	\\
J1446-4701	&	17	&	68	&	51	&	80	&	25	&	-55	\\
J1514-4946	&	24	&	68	&	44	&	25	&	68	&	43	\\
J1600-3053	&	61	&	37	&	-24	&	65	&	28	&	-37	\\
J1614-2230	&	80	&	80	&	0	&	64	&	88	&	24	\\
J1658-5324	&	30	&	69	&	39	&	78	&	23	&	-55	\\
J1713+0747	&	36	&	68	&	32	&	36	&	65	&	29	\\
J1747-4036	&	19	&	80	&	61	&	19	&	80	&	61	\\
J2017+0603	&	34	&	67	&	33	&	23	&	74	&	51	\\
J2043+1711	&	54	&	76	&	22	&	53	&	79	&	26	\\
J2047+1053	&	51	&	71	&	20	&	51	&	71	&	20	\\
J2051-0827	&	43	&	69	&	26	&	68	&	42	&	-26	\\
J2215+5135	&	18	&	71	&	53	&	18	&	71	&	53	\\
J2241-5236	&	20	&	76	&	56	&	19	&	75	&	56	\\
J2302+4442	&	60	&	46	&	-14	&	65	&	37	&	-28	\\

     	\hline

	\end{tabular}
    	\caption{Magnetic impact angle ($\beta$) for Class I MSPs determined from inclination angles in \protect\cite{RN28} provided for a two-pole caustic model (TPC), and an outer-gap (OG) model.
    	} 
        \label{tab:RN28_TABLE_6_AND_7}
\end{table*}

\begin{table*}
 \centering
	
	\begin{tabular}{lc c c c c c c c c c } 
		\hline
	&		&	TPC 	&		&		&	OG	&		&		&	SG	&		\\
MSP	&	$\alpha$ (\textdegree)&	$\zeta$	(\textdegree)&	$\beta$	(\textdegree)&	$\alpha$ (\textdegree)&	$\zeta$	(\textdegree)&	$\beta$	(\textdegree)&	$\alpha$ (\textdegree)	&	$\zeta$	(\textdegree)&	$\beta$	(\textdegree)\\

        \hline

J0034-0534	&	23	&	69	&	46	&	22	&	75	&	53	&	23	&	74	&	51	\\
J1810+1744	&	82	&	16	&	-66	&	81	&	24	&	-57	&	5	&	29	&	24	\\
J1823-3021A	&	46	&	52	&	6	&	42	&	78	&	36	&	78	&	45	&	-33	\\
J1902-5105	&	10	&	73	&	63	&	10	&	73	&	63	&	56	&	16	&	-40	\\
J1939+2134	&	88	&	88	&	0	&	72	&	85	&	13	&	33	&	41	&	8	\\
J1959+2048	&	56	&	85	&	29	&	52	&	87	&	35	&	46	&	35	&	-11	\\

        \hline

	\end{tabular}
    	\caption{Magnetic impact angle ($\beta$) for Class II MSPs determined from inclination angles in \protect\cite{RN28} provided for a two-pole caustic  (TPC), an outer-gap (OG) and a slot-gap (SG) model.
    	} 
        \label{tab:RN28_TABLE_8_AND_9_AND_10}
\end{table*}

\begin{table}
 \centering
	
	\begin{tabular}{lc c c c } 
		\hline
  
MSP	&	$\alpha$ (\textdegree)&	$\zeta$	(\textdegree)&	$\beta$	(\textdegree)\\
\hline
J0340+4130	&	43	&	73	&	30	\\
J1741+1351	&	46	&	80	&	34	\\
J1744-1134	&	51	&	85	&	34	\\
J1858-2216	&	42	&	74	&	32	\\
J2124-3358	&	19	&	25	&	6	\\
J2214+3000	&	59	&	86	&	27	\\

        \hline

	\end{tabular}
    	\caption{Magnetic impact angle ($\beta$) for Class III MSPs determined for inclination angles in \protect\cite{RN28} provided for a pair-starved polar cap (PSPC) model.
    	} 
        \label{tab:RN28_TABLE_11}
\end{table}

\begin{table}
 \centering
	
	\begin{tabular}{lc c c c } 
		\hline

MSP	&	$\alpha$ (\textdegree)	&	$\zeta$	(\textdegree)&	$\beta$	(\textdegree)\\
\hline
 J0030+0451	&	70	&	60	&	-10	\\
 J0102+4839	&	55	&	70	&	15	\\
 J0437-4715	&	45	&	40	&	-5	\\
 J0614-3329	&	75	&	56	&	-19	\\
 J1124-3653	&	65	&	46	&	-19	\\
 J1514-4946	&	55	&	62	&	7	\\
 J1614-2230	&	85	&	86	&	1	\\
 J2017+0603	&	55	&	48	&	-7	\\
 J2043+1711	&	55	&	74	&	19	\\
 J2302+4442	&	50	&	60	&	10	\\

        \hline

	\end{tabular}
    	\caption{Magnetic impact angle ($\beta$) for Class I MSPs determined from inclination angles provided in \protect\cite{RN635}.
    	} 
        \label{tab:RN635_TABLE_1}
\end{table}

More recently, \cite{RN635} determine $\zeta$ and $\alpha$ for 10 of the same Class I MSPs as \cite{RN28}, using a force-free magnetospheric model for gamma-ray emission at the light cylinder and into the striped wind, and radio emission from the polar-cap to reproduce gamma-ray and radio light curves from the 2PC. They then fit the time-lag between radio and first gamma-ray pulse along with the gamma-ray pulse separation and ratio of the gamma-ray pulse heights of light curves in the 2PC for varying inclination and magnetic angle to obtain a best fit. We again determine $\beta$ from their inclination and magnetic angles (Table ~\ref{tab:RN635_TABLE_1}).

\begin{figure}
\includegraphics[width=0.5\textwidth]{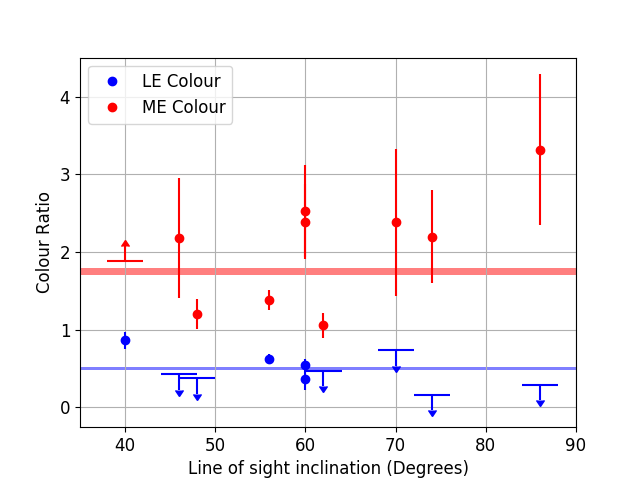}
\\
\includegraphics[width=0.5\textwidth]{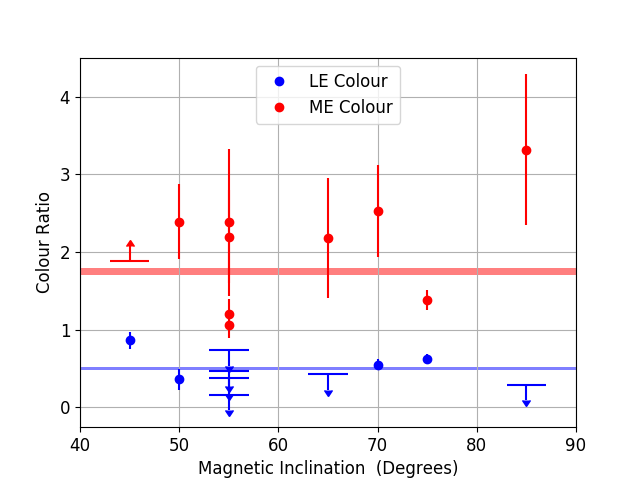}
\quad
\includegraphics[width=0.5\textwidth]{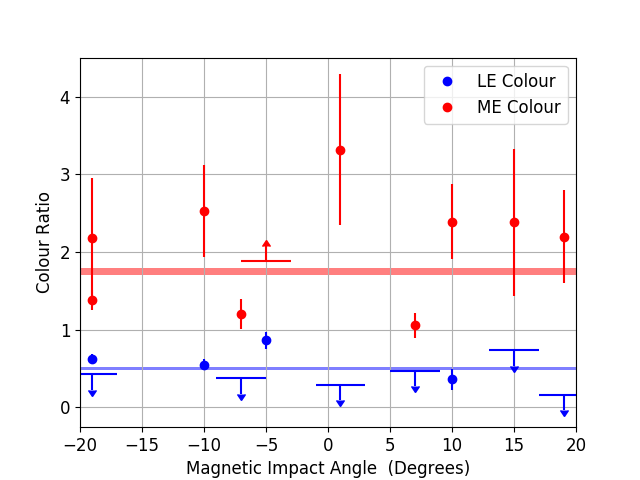} 
\\
    \caption{The LE and ME colours for ten MSPs plotted against line of sight inclination ($\zeta$), magnetic inclination ($\alpha$) from \protect\cite{RN635} and magnetic impact angle ($\beta=$Line of sight inclination $-$  magnetic inclination). All inclinations are with respect to the pulsar rotational axis. There is no obvious correlation of colour with line of sight or magnetic inclination whilst the magnetic impact angle shows colours with symmetry between 5 and 10\textdegree{} where the colour ratios are close to 1 indicating a harder spectrum. The LP colour values are shown as red and blue bands. Colour ratios are shown either as points with errors, or as upper or lower limits.}
\label{fig:colour_inclinations}   
\end{figure}

We then plot $\zeta$, $\alpha$ and $\beta$ for MSPs in our selection in Fig.~\ref{fig:colour_inclinations}.  The values of $\zeta$ and $\alpha$ are uncorrelated with the model colours we have calculated. However, for $\beta$, there is some indication of harder ME emission for absolute $\beta$ values of  5$-$10\textdegree{} (ME colour ratio $\sim$1) and generally softer emission outside this range, (ME colour ratio $>$ 2), suggesting that ME spectral features are influenced by the offset of the emission region from the pulsar spin axis, once geometric viewing effects are allowed for. However, the inclination angles of more MSPs would need to be determined with the same force-free model to draw a definitive conclusion, especially as no such effect is seen when using the inclination angles of \cite{RN28}. 

In conclusion, it appears that pulsar line-of-sight and the magnetic axis inclination seem to have little effect on the observed gamma-ray spectrum at low and medium energies. Therefore the observed gamma-ray spectrum does not allow us to constrain pulsar geometry or to further validate the existing synchro-curvature models of pulsar emission which incorporate geometric effects. However, $\zeta$ and $\alpha$ are determined only for a relatively small pulsar population and so we cannot completely exclude spectral differences arising from pulsar geometry. 

Our standalone LP model is also useful in its own right to model ensemble emission from a population of MSPs. Such an application of a stacked MSP model is demonstrated in \cite{RN430}, wherein, the gamma-ray emission of the globular cluster 47 Tuc is modeled using the MSP model of \cite{RN183} combined with an annihilating dark matter (DM) model to show that a "MSP+DM" model is preferred over MSP emission alone. In this regard we expect our LP model comprising a larger sample of 118 MSPs with well constrained 1$\sigma$ errors (Fig.~\ref{fig:replace_bands}) to be more representative of an MSP ensemble than that of \cite{RN183} which uses just 39 MSPs, a much greater 3$\sigma$ uncertainty and only 57~\% of the photon exposure of our study. The caveat applies that any stacked spectrum should be applied to systems only where the total spectral variation of individual MSPs is less than the model hypothesis under consideration, as demonstrated by the comment on \cite{RN430} in \cite{RN739} and its reply in \cite{RN740}. Specifically, our model will be applicable to considering the ensemble gamma-ray emission from MSPs in globular clusters (GCs, \cite{RN196}), as the characteristic ages\footnote{Calculated for the list of MSPs in GCs (with period and positive period derivative)  maintained at http://www.naic.edu/{\textasciitilde}pfreire/GCpsr.html accessed 29/10/2022} of 76 GC MSPs ($2.6 \times 10\textsuperscript{7}-6.8 \times 10\textsuperscript{10}$ yrs) are consistent with those of the model MSPs in Fig.~\ref{fig:colour_msp_key_props}. Furthermore, the spectra of disc and bulge MSPs are consistent within uncertainties (\cite{Ploeg_2020} ) and so we expect our model based on local disc MSPs to be useful in consideration of the Galactic centre excess problem which could arise from an unresolved population of MSPs (\cite{RN78}).

\section{Conclusion}
\label{sec:Conclusion}

We analyse 127 MSPs from the Public List of LAT Detected Gamma-Ray Pulsars and detect 118 MSPs in the range 100 MeV$-$100 GeV.  We sum the 100 MeV$-$56.2 GeV fluxes of these MSPs and find the best fit to the resulting spectral energy distribution is an LP model and that this model is superior to other published models of stacked MSP emission. 

Most previous models of gamma-ray emission from MSPs suggest that pulsar geometry may affect the observed gamma-ray spectrum. These models are difficult to compare with observations as they are either qualitative, do not cover the full range of geometries or are a proof-of-concept of modelling technique which does not reproduce known features of MSP spectra. 

In the absence of quantitative spectral emission models for comparison, we use our LP model as a baseline to determine if pulsar properties affect spectral shape by using energy flux ratios (or colours) for the low-energy (133/237 MeV) and medium energy (4.2/7.5 GeV) bins. We find that pulsar $\dot{E}$, surface magnetic field and gamma-ray luminosity are uncorrelated with these colours and hence do not systematically affect the spectral shape in the low-energy and medium energy bins. Similarly we find that LE and ME colours, are in the main, uncorrelated with magnetic impact angle, indicating that pulsar spin axis inclination and magnetic inclination have little effect on the LE and ME features of the observed gamma-ray spectrum. 
There is a hint of symmetry in ME colours for harder emission from  magnetic impact angles between 5 and 10 \textdegree, but more inclination determinations for a larger pulsar sample would be required to confirm this. 

In general, our results show little influence on the low and medium energy gamma-ray spectrum arising from pulsar geometry.

Finally, we note that our MSP spectral model (LP) is useful more generally in the problems of the Galactic centre excess and the ensemble emission in globular clusters arising from a population of MSPs.

\section*{Acknowledgements}

We acknowledge the data and tools provided by the \textit{Fermi}-LAT collaboration. AMB and PMC acknowledge the financial support of the UK Science and Technology Facilities Council consolidated grant ST/X001075/1. SJL acknowledges the financial support of Durham University. We thank the anonymous referee for comments which helped improve this paper.

\section*{Data Availability Statement}
The data underlying this article will be shared on reasonable request to the corresponding author.

\appendix

\section{Comparison of Stacked Models of MSP Emission}
\label{sec:DiscussionModels}

The 2PC is a published survey of pulsars observed by \textit{Fermi}-LAT. It uses 3 years of  \textsc{pass} 7 event data in the energy range between 100 MeV and 100 GeV with the 2FGL source catalog as a source model and lists the spectral models and fluxes of 117 pulsars, evenly divided between MSPs, young radio-quiet and young radio-loud pulsars. The survey uses three search strategies for pulsar detection to overcome the difficulty of only one photon being detected in a few million pulsar rotations. Firstly, the known rotation ephemerides of pulsars, obtained mostly through radio and in some cases X-ray observations, are used to fit a timing model with \textsc{tempo/tempo2} software, to tag the gamma-ray event data with a pulsar phase. The gamma-ray data are then phase-folded to identify any  emission peaks. Secondly, blind periodicity searches are used on unassociated sources classed as candidate pulsars because they show no variability and have spectra that can be fitted by an exponential cut-off in the GeV band. This method is challenging because the event data are sparse with only a few photons detected per hour and in addition pulsars may have been missed by being in binary systems, which tends to smear the signal through Doppler shifts arising from orbital motion. Finally, the detection of pulsed radio emission in unassociated sources and the construction of timing models can lead to the detection of gamma-ray pulsations through phase folding methods as above. The 2PC lists 40 MSPs, 20 of which have been detected using this final method. The 2PC increased the then known MSP sample from 8 to 40 MSPs with heliocentric distances up to 2 kpc and a uniform distribution in the sky. The MSPs exhibit between 1$-$3 gamma-ray peaks and their differential flux spectrum, \textit{dN/dE}, (photon flux per energy bin) is an exponential cut-off power law as described by Eqn.~\ref{CH5PLexp}. This equation is functionally equivalent to Eqn.~\ref{PLexp} with the exponential factor $a$ replaced by $1/E\textsubscript{cut}$ and index $\gamma_1$ having negative sign. For consistency with the 2PC and \cite{RN183} we continue to use symbols $k$,  $\Gamma$ and $b$ which are equivalent to the $N\textsubscript{0}$, $\gamma_1$ and $\gamma_2$ in  Eqn.~\ref{PLexp}

\begin{equation}{\label{CH5PLexp}}
\frac{dN}{dE}=k\Big(\frac{E}{E\textsubscript{0}}\Big)\textsuperscript{-$\Gamma$}\exp\Big(-\frac{E}{E\textsubscript{cut}}\Big)\textsuperscript{b}
\end{equation}

\begin{equation}{\label{McCann2016}}
E^2\frac{dN}{dE}=k\Big(\frac{E}{1 GeV}\Big)\textsuperscript{$\Gamma$}\exp\Big(-\frac{E}{E\textsubscript{cut}}\Big)\textsuperscript{b}
\end{equation}

 The MSPs listed in the 2PC also provide a pulsar sample to use in determining models of stacked gamma-ray emission.  \cite{RN183} re-analyse 39 of the 40 MSPs in the 2PC (excluding one, J1939+2134, which has a detection significance of $\simeq 3\sigma$ ) using 7.5 years of \textsc{pass 8} event data in the energy range 100 MeV$-$300 GeV in 15 energy bins, using the 3FGL as a source catalog model. 
 
 In the 3FGL, 33 MSPs are described by an exponential cut-off power law model (Eqn.~\ref{CH5PLexp}), whereas 6 MSPs are best fitted with a simple power law (PL) (Eqn.~\ref{PLeqn}). However, in their analysis they find that an exponential cut-off can be detected in the 6 PL MSPs at > 3 $\sigma$ significance. They therefore use an exponential cut-off power law model throughout their analysis. They then stack all flux points from the 39 MSPs with a TS > 9 (equivalent to > 3 $\sigma$ significance) to obtain a functional form described by an exponential cut-off power law as Eqn.~\ref{CH5PLexp} with $\Gamma = 1.54\substack{+0.10 \\ -0.11}$ and $E_{cut} = 3.70\substack{+0.95 \\ -0.70}$ but with $E_0$ and $b$ equal to 1 (hereafter the "Xing and Wang" model). Finally, they recommend that this functional form can be used as a model to find candidate MSPs in unidentified \textit{Fermi}-LAT sources at high Galactic latitudes.

Alternatively, \cite{RN197} constructs a stacked MSP gamma-ray spectrum using an aperture photometry (AP) method rather than likelihood analysis. The AP approach has the advantage over the likelihood approach in that it is model independent and less computationally intensive. However, it does require timing information for the pulsars analysed. McCann chooses 39 MSPs from the 2PC (excluding a different one, J2215+5135 because its off-phase, where emission is at a minimum, is undefined) and consider 4.2 years of \textsc{pass 7} event data per MSP in the energy range 100 MeV$-$1 TeV binned at 4 bins per decade of energy. McCann then uses the \textsc{tempo 2} software to barycentre and phase fold the photon events. McCann then obtains the energy excess counts of all events outside the off-phase (i.e the on-phase), distributed by energy, corrects for exposure and produces a spectral energy distribution from the stacked fluxes. Finally McCann fits the differential flux $E^2\frac{dN}{dE}$ (as opposed $\frac{dN}{dE}$)  with an exponential cut-off power law, with a functional form as Eqn.~\ref{McCann2016}. This exponential cut-off power law has  $\Gamma = 0.54\pm0.05$, $E\textsubscript{cut} = 3.60\pm0.21$ GeV and $b=0.7\pm0.15$ (hereafter the "McCann" model). McCann also makes a check on the performance of the AP method vs the likelihood method of the 2PC by defining a flux ratio for the MSPs of $\frac{AP flux}{2PC flux}$ which varies between 0.8 and 0.9 for energies of 250 MeV$-$8 GeV. 
 \begin{table}
 \centering
	
	\begin{tabular}{lc c c} 
		\hline
                Parameter	& Value & Unit \\
		\hline

\textit{Index1}&  -1.28$\ \pm \ $0.01 & - \\
\textit{Normalisation}& (4.75$\ \pm \ {0.06} \ )\ \times \ $10\textsuperscript{-4}  & cm\textsuperscript{-2} s\textsuperscript{-1} TeV\textsuperscript{-1} \\
\textit{Exponential factor}& (3.95$\ \pm \ {0.07} \ )  \ \times \  $10\textsuperscript{2} & TeV \textsuperscript{-1} \\
\textit{Index2}& 1.0 & - \\
\textit{Scale}&  1$\ \times \ $10\textsuperscript{-3} & TeV  \\

		\hline
	\end{tabular}
    	\caption{The parameters of the best-fitted PLSEC model (this work) using Eqn.~\ref{PLexp}, for the stacked differential energy flux of 118 significantly detected MSPs in the energy range 100 MeV$-$56.2 GeV. This model is inferior to the LP model above.}
        \label{tab:PLEC_MODEL_PARAMS}
\end{table}

We assess which MSP model(s) (LP, PLSEC, McCann or Xing and Wang) are a preferred description of our stacked MSP spectrum through a likelihood analysis. We allow the normalisation of the models to vary and determine for each normalisation the residual for each energy bin which is the difference between the stacked MSP flux and the model flux evaluated at the centre of each energy bin in the spectrum. We sum the log of the absolute value of each bin residual to obtain a set of log likelihood values, one for each normalisation. We then determine a minimum log likelihood value of $-135.2$, $-131.9$, $-137.6$ and $-123.9$ for the LP, PLSEC, Xing and Wang, and McCann models, respectively. Using Eqn.~\ref{MATTOX_TS} and the minimum log likelihood value of each model as $L_0$ and $L_1$, we determine that LP, and Xing and Wang, are significantly preferred over that of McCann at TS 22.6 (4.7$\sigma$) and 27.4 (5.2$\sigma$) respectively, whilst LP, and Xing and Wang are an equally good fit to the MSP spectrum using a likelihood analysis.

In Fig.~\ref{fig:replace_bands}, we show the likelihood best fit  models and the stacked spectrum of 118 MSPs with the LP model fitting the greatest number of flux points. 

The preferred MSP model can also be determined using the minimum value of the Akaike Information Criterion (AIC) statistic \cite{AIC}, (Eqn.~\ref{AIClikelihood}). The AIC ranks how well a model fits a data set (compared to other models) and penalises the over-fitting which results from the model having more free parameters. The AIC is a relative measure in that it allows a set of models to be compared with the model exhibiting the lowest AIC score considered superior in that set, but it does not allow a determination of whether any model is best in an absolute sense. The AIC is defined in Eqn.~\ref{AIClikelihood} where $k$ is the number of free model parameters and $\ell$ is the likelihood of the best fit model.      

\begin{equation}{\label{AIClikelihood}}
AIC=2k-2\ln(\ell)
\end{equation}

For the purposes of spectral model comparison, a more convenient definition of AIC is Eqn.~\ref{AICRSS} where $n$ is the number of flux data points or energy bins and $RSS$ is the residual sum of squares as defined in Eqn.~\ref{RSS} with $y_i$ being the observed flux and $f(x_i)$  the flux predicted by the model for at an energy $x_i$ for an energy bin $i$. 

\begin{equation}{\label{AICRSS}}
AIC=2k+n\ln\big(\frac{RSS}{n}\big)
\end{equation}

\begin{equation}{\label{RSS}}
RSS=\sum_{i=1}^{n} (y_i-f(x_i))^2
\end{equation}

The AIC statistic for the fit of our models, LP, PLSEC, and the models of Xing and Wang, and McCann, to the stacked MSP spectrum between 100 MeV$-$56.2 GeV, is $-232.7$, $-223.9$, $-223.1$ and $-204.4$ respectively. Our model, LP, having the minimum AIC statistic, is thus the preferred one. Furthermore the evidential significance $\Delta_i$ of any model, $i$, can be determined as $\Delta_i$ = (AIC value for model $i$ - AIC value for our model). A $\Delta_i$ value $\leq 2$ indicates a model has substantial support whereas $\Delta_i \geq 10$ indicates a model has essentially no support (\cite{RN741}).  The $\Delta_i$ values for Xing and Wang, and McCann are 9.6 and 28.3 respectively and hence these models are disfavoured by the AIC statistic.

The absolute goodness-of-fit of models to the stacked MSPs can also be determined by the $\chi^2$ statistic which we calculate across the energy range of each bin and take the minimum (best) $\chi^2$ value in each bin for the best fit likelihood model. 
The LP, PLSEC, Xing and Wang and McCann models have a $\chi^2$ statistic of 6.6, 192.0, 1225.5 and 2472.8 respectively between 100 MeV$-$56.2 GeV (Table~\ref{tab:CH5_CHI_SQUARE}).

The Xing and Wang model is an acceptable fit from 178 MeV to 56.2 GeV whereas the PLSEC model is an acceptable fit from 562 MeV to 17.8 GeV. The McCann model is not a good fit, with $\chi^2$ values exceeding the critical value in most bins (7 d.o.f. / critical value 24.3 for $\alpha = 0.001$). In contrast, only the LP model provides an acceptable fit across the whole energy range (8 d.o.f. / critical value 26.1 for $\alpha = 0.001$) whilst minimising $\chi^2$ compared to the other models,
The LP model is therefore the preferred spectral model overall.

During the final preparation of this paper we became aware of a new MSP spectral model based on 104 MSPs (\cite{Wu_2022}) prepared using the same method as \cite{RN183}. However, we find that this model is essentially the same as that of \cite{RN183}, having log-likelihood $=$ -142.1 and AIC $=$ -221.5, and similar $\chi^2$ per bin with respect to the critical value as \cite{RN183}, so we do not consider it further.


\begin{table*}
 \centering
	
	\begin{tabular}{lc c c c c c c } 
		\hline
Bin Center	& Lower Bin Energy &Upper Bin Energy&	LP           &  PLSEC         &  Xing and Wang	&  McCann 	\\
MeV         & MeV              & MeV            &	  $\chi^2$   & $\chi^2$       &  $\chi^2$         &  $\chi^2$  \\        
            
133	    &	100	    &	178	    &	0.0	&	121.1	&	1198.1	&	199.9	\\
237	    &	178	    &	316	    &	0.0	&	15.6	&	0.0	    &	101.4	\\
421	    &	316	    &	562	    &	0.0	&	27.8	&	0.0	    &	445.3	\\
750	    &	562	    &	1000	&	0.0	&	0.0	    &	1.6	    &	631.9	\\
1333	&	1000	&	1778	&	6.6	&	0.0	    &	20.8	&	638.7	\\
2371	&	1778	&	3162	&	0.0	&	0.0	    &	0.0	    &	210.4	\\
4217	&	3162	&	5623	&	0.0	&	0.0	    &	0.0	    &	0.0	\\
7499	&	5623	&	10000	&	0.0	&	0.0	    &	0.0	    &	65.5	\\
13335	&	10000	&	17783	&	0.0	&	0.0	    &	0.0	    &	157.4	\\
23714	&	17783	&	31623	&	0.0	&	21.5	&	0.1	    &	22.3	\\
42170	&	31623	&	56234	&	0.0	&	6.0	    &	4.9	    &	0.0	\\

        \hline
        Total:&&& 6.6	&	192.0	&	1225.5 &		2472.8 \\

     	\hline

	\end{tabular}
    	\caption{A breakdown of $\chi^2$ test statistic by energy bin, for stacked models of MSP emission fitted to the 118 MSP stacked flux, ranked in order of total $\chi^2$. The LP  model (this work) is best in minimising $\chi^2$ overall and the only model which provides an acceptable fit across the whole energy range (0.1$-$56.2 GeV). The McCann model is not a good fit, with $\chi^2$ values exceeding the critical value in most bins (7 d.o.f. / critical value 24.3 for $\alpha = 0.001$). All other models 8 d.o.f. / critical value 26.1 for $\alpha = 0.001$. 
    	} 
        \label{tab:CH5_CHI_SQUARE}
\end{table*}

\onecolumn
\section{MSP fluxes and detection significance}

\begin{longtable}{lc c c c c c }

	\hline
	MSP Name	&	Source ID	&	TS	&	Offset	&	Energy Flux			&	Photon Flux			\\		           	& 	&	 	&	(Degrees)	&	10\textsuperscript{-11} erg cm\textsuperscript{-2} s\textsuperscript{-1}			&	10\textsuperscript{-8} cm\textsuperscript{-2} s\textsuperscript{-1}			\\	        
    \hline
    \endfirsthead
    \multicolumn{3}{@{}l}{\ldots continued}\\\hline
	MSP Name	&	Source ID	&	TS	&	Offset	&	Energy Flux			&	Photon Flux			\\		           	& 	&	 	&	(Degrees)	&	10\textsuperscript{-11} erg cm\textsuperscript{-2} s\textsuperscript{-1}			&	10\textsuperscript{-8} cm\textsuperscript{-2} s\textsuperscript{-1}			\\	        

    \hline
    \endhead
    \hline
    \multicolumn{3}{r@{}}{continued \ldots}\\
    \endfoot
    \hline
    \endlastfoot

J1921+1929 & 4FGL J1921.1+1930 & 27.5  & 0.058  & 0.44 $\pm$ 0.32  & 0.52 $\pm$ 0.21 \\
J0737-3039A & PS J0738.0-3041 & 29.3  & 0.056  & 0.34 $\pm$ 0.09  & 0.30 $\pm$ 0.12 \\
J1137+7528 & 4FGL J1137.6+7527 & 29.4  & 0.043  & 0.07 $\pm$ 0.23  & 0.03 $\pm$ 0.10 \\
J1125-6014 & 4FGL J1126.4-6011 & 33.6  & 0.086  & 0.28 $\pm$ 0.82  & 0.13 $\pm$ 0.25 \\
J1811-2405 & 4FGL J1811.3-2403 & 37.3  & 0.035  & 0.93 $\pm$ 0.18  & 1.34 $\pm$ 0.36 \\
J1833-3840 & 4FGL J1833.0-3840 & 54.7  & 0.012  & 0.24 $\pm$ 0.51  & 0.19 $\pm$ 0.07 \\
J0931-1902 & 4FGL J0931.2-1906 & 57.5  & 0.066  & 0.16 $\pm$ 0.45  & 0.10 $\pm$ 0.29 \\
J0653+4706 & 4FGL J0652.9+4707 & 63.4  & 0.028  & 0.17 $\pm$ 0.04  & 0.16 $\pm$ 0.06 \\
J0248+4230 & 4FGL J0248.6+4230 & 75.7  & 0.021  & 0.19 $\pm$ 0.34  & 0.17 $\pm$ 0.11 \\
J2052+1218 & 4FGL J2052.7+1218 & 75.9  & 0.015  & 0.49 $\pm$ 0.07  & 0.76 $\pm$ 0.16 \\
J1431-4715 & 4FGL J1431.4-4711 & 79.0  & 0.080  & 0.54 $\pm$ 0.07  & 1.00 $\pm$ 0.18 \\
J1855-1436 & 4FGL J1855.9-1435 & 83.2  & 0.017  & 0.50 $\pm$ 1.04  & 0.37 $\pm$ 0.10 \\
J0952-0607 & 4FGL J0952.1-0607 & 83.9  & 0.020  & 0.22 $\pm$ 0.43  & 0.16 $\pm$ 0.07 \\
J1730-2304 & 4FGL J1730.8-2303 & 87.8  & 0.106  & 0.76 $\pm$ 0.61  & 1.05 $\pm$ 0.16 \\
J1908+2105 & 4FGL J1908.9+2103 & 90.2  & 0.020  & 0.54 $\pm$ 0.22  & 0.55 $\pm$ 0.07 \\
J2051-0827 & 4FGL J2051.0-0826 & 91.4  & 0.028  & 0.25 $\pm$ 0.56  & 0.12 $\pm$ 0.28 \\
J1555-2908 & 4FGL J1555.7-2908 & 92.8  & 0.006  & 0.57 $\pm$ 0.07  & 0.96 $\pm$ 0.15 \\
J1640+2224 & 4FGL J1640.1+2222 & 103.5  & 0.047  & 0.27 $\pm$ 0.36  & 0.29 $\pm$ 0.08 \\
J1544+4937 & 4FGL J1544.0+4939 & 105.9  & 0.022  & 0.23 $\pm$ 0.06  & 0.22 $\pm$ 0.06 \\
J0621+2514 & 4FGL J0621.2+2512 & 115.3  & 0.026  & 0.48 $\pm$ 0.51  & 0.30 $\pm$ 0.04 \\
J2006+0148 & 4FGL J2006.4+0147 & 117.5  & 0.022  & 0.39 $\pm$ 0.14  & 0.21 $\pm$ 0.06 \\
J2039-3616 & 4FGL J2039.4-3616 & 120.1  & 0.029  & 0.37 $\pm$ 0.50  & 0.47 $\pm$ 0.15 \\
J1641+8049 & 4FGL J1641.2+8049 & 132.5  & 0.006  & 0.22 $\pm$ 0.08  & 0.23 $\pm$ 0.05 \\
J2047+1053 & 4FGL J2047.3+1051 & 133.0  & 0.058  & 0.43 $\pm$ 0.44  & 0.32 $\pm$ 0.02 \\
J1732-5049 & 4FGL J1732.7-5050 & 144.7  & 0.021  & 0.56 $\pm$ 0.68  & 0.62 $\pm$ 0.13 \\
J1741+1351 & 4FGL J1741.4+1354 & 147.0  & 0.051  & 0.43 $\pm$ 0.06  & 0.31 $\pm$ 0.08 \\
J1125-5825 & 4FGL J1125.6-5825 & 149.7  & 0.012  & 0.59 $\pm$ 0.71  & 0.37 $\pm$ 0.09 \\
J1805+0615 & 4FGL J1805.6+0615 & 165.3  & 0.007  & 0.53 $\pm$ 0.79  & 0.36 $\pm$ 0.11 \\
J1939+2134 & 4FGL J1939.6+2135 & 178.0  & 0.016  & 1.97 $\pm$ 0.33  & 3.04 $\pm$ 0.43 \\
J1552+5437 & 4FGL J1553.1+5438 & 181.4  & 0.042  & 0.28 $\pm$ 0.26  & 0.23 $\pm$ 0.05 \\
J1824+1014 & 4FGL J1824.1+1013 & 185.1  & 0.021  & 0.59 $\pm$ 0.71  & 0.44 $\pm$ 0.04 \\
J0740+6620 & 4FGL J0741.0+6618 & 192.5  & 0.038  & 0.30 $\pm$ 1.09  & 0.28 $\pm$ 0.10 \\
J1036-8317 & 4FGL J1036.5-8318 & 195.7  & 0.013  & 0.44 $\pm$ 0.05  & 0.30 $\pm$ 0.06 \\
J1024-0719 & 4FGL J1024.5-0719 & 200.6  & 0.034  & 0.43 $\pm$ 0.54  & 0.44 $\pm$ 0.12 \\
J2115+5448 & 4FGL J2115.1+5449 & 205.9  & 0.014  & 0.80 $\pm$ 0.28  & 0.48 $\pm$ 0.03 \\
J1335-5656 & 4FGL J1335.0-5656 & 211.8  & 0.017  & 0.87 $\pm$ 0.43  & 0.71 $\pm$ 0.31 \\
J1012-4235 & 4FGL J1012.1-4235 & 218.5  & 0.011  & 0.52 $\pm$ 0.05  & 0.50 $\pm$ 0.09 \\
J1921+0137 & 4FGL J1921.4+0136 & 226.5  & 0.027  & 1.04 $\pm$ 0.22  & 1.07 $\pm$ 0.18 \\
J1207-5050 & 4FGL J1207.4-5050 & 232.7  & 0.012  & 0.53 $\pm$ 0.06  & 0.44 $\pm$ 0.11 \\
J0955-6150 & 4FGL J0955.4-6151 & 238.0  & 0.024  & 0.69 $\pm$ 0.55  & 0.77 $\pm$ 0.06 \\
J0251+2606 & 4FGL J0251.0+2605 & 239.0  & 0.007  & 0.50 $\pm$ 0.43  & 0.47 $\pm$ 0.04 \\
J1446-4701 & 4FGL J1446.6-4701 & 240.2  & 0.008  & 0.63 $\pm$ 0.73  & 0.57 $\pm$ 0.12 \\
J1827-0849 & 4FGL J1827.6-0849 & 243.9  & 0.008  & 2.31 $\pm$ 1.85  & 2.50 $\pm$ 0.58 \\
J1048+2339 & 4FGL J1048.6+2340 & 250.0  & 0.021  & 0.56 $\pm$ 0.05  & 0.93 $\pm$ 0.14 \\
J2042+0246 & 4FGL J2042.2+0245 & 263.6  & 0.074  & 0.62 $\pm$ 0.37  & 0.80 $\pm$ 0.05 \\
J1543-5149 & 4FGL J1543.6-5148 & 270.8  & 0.016  & 1.66 $\pm$ 1.09  & 2.43 $\pm$ 0.31 \\
J2017-1614 & 4FGL J2017.7-1612 & 271.1  & 0.036  & 0.65 $\pm$ 0.49  & 0.69 $\pm$ 0.18 \\
J0318+0253 & 4FGL J0318.2+0254 & 279.1  & 0.028  & 0.59 $\pm$ 0.60  & 0.58 $\pm$ 0.15 \\
J1843-1113 & 4FGL J1843.7-1114 & 282.1  & 0.021  & 1.71 $\pm$ 1.16  & 3.05 $\pm$ 0.47 \\
J1400-1431 & 4FGL J1400.6-1432 & 294.6  & 0.019  & 0.66 $\pm$ 0.07  & 0.77 $\pm$ 0.15 \\
J1649-3012 & 4FGL J1649.8-3010 & 302.6  & 0.032  & 0.97 $\pm$ 0.73  & 1.10 $\pm$ 0.23 \\
J1622-0315 & 4FGL J1623.0-0315 & 321.7  & 0.004  & 0.91 $\pm$ 0.66  & 1.10 $\pm$ 0.25 \\
J0605+3757 & 4FGL J0605.1+3757 & 326.4  & 0.009  & 0.67 $\pm$ 0.07  & 0.42 $\pm$ 0.04 \\
J0312-0921 & 4FGL J0312.1-0921 & 327.3  & 0.016  & 0.51 $\pm$ 0.05  & 0.31 $\pm$ 0.05 \\
J2129-0429 & 4FGL J2129.8-0428 & 328.5  & 0.026  & 0.58 $\pm$ 0.69  & 0.51 $\pm$ 0.14 \\
J1713+0747 & 4FGL J1713.8+0747 & 331.7  & 0.010  & 0.77 $\pm$ 0.07  & 0.65 $\pm$ 0.11 \\
J1513-2550 & 4FGL J1513.4-2549 & 351.4  & 0.024  & 0.75 $\pm$ 0.69  & 0.76 $\pm$ 0.15 \\
J1221-0633 & 4FGL J1221.4-0634 & 356.4  & 0.031  & 0.68 $\pm$ 0.70  & 0.76 $\pm$ 0.16 \\
J1903-7051 & 4FGL J1903.4-7053 & 360.0  & 0.033  & 0.58 $\pm$ 0.06  & 0.48 $\pm$ 0.09 \\
J1628-3205 & 4FGL J1628.1-3204 & 374.1  & 0.026  & 1.14 $\pm$ 0.57  & 1.56 $\pm$ 0.29 \\
J1142+0119 & 4FGL J1142.8+0120 & 385.2  & 0.011  & 0.66 $\pm$ 0.07  & 0.57 $\pm$ 0.11 \\
J1901-0125 & 4FGL J1901.4-0126 & 394.1  & 0.028  & 1.91 $\pm$ 1.31  & 2.52 $\pm$ 0.28 \\
J2310-0555 & 4FGL J2310.0-0555 & 402.4  & 0.013  & 0.76 $\pm$ 0.07  & 0.96 $\pm$ 0.16 \\
J1747-4036 & 4FGL J1747.7-4037 & 410.9  & 0.004  & 1.28 $\pm$ 0.79  & 1.67 $\pm$ 0.21 \\
J0610-2100 & 4FGL J0610.2-2100 & 411.4  & 0.010  & 0.68 $\pm$ 0.40  & 0.72 $\pm$ 0.28 \\
J1600-3053 & 4FGL J1600.9-3054 & 422.0  & 0.012  & 0.85 $\pm$ 0.69  & 0.47 $\pm$ 0.03 \\
J0023+0923 & 4FGL J0023.4+0920 & 426.5  & 0.053  & 0.82 $\pm$ 0.06  & 0.94 $\pm$ 0.13 \\
J1745+1017 & 4FGL J1745.5+1017 & 432.9  & 0.011  & 0.88 $\pm$ 0.78  & 0.72 $\pm$ 0.14 \\
J1630+3734 & 4FGL J1630.6+3734 & 441.3  & 0.010  & 0.55 $\pm$ 0.52  & 0.42 $\pm$ 0.10 \\
J2034+3632 & 4FGL J2035.0+3632 & 483.2  & 0.006  & 1.26 $\pm$ 0.77  & 0.40 $\pm$ 0.12 \\
J1301+0833 & 4FGL J1301.6+0834 & 496.0  & 0.005  & 0.79 $\pm$ 0.27  & 0.83 $\pm$ 0.12 \\
J2256-1024 & 4FGL J2256.8-1024 & 520.7  & 0.014  & 0.81 $\pm$ 0.72  & 0.92 $\pm$ 0.07 \\
J1823-3021A & 4FGL J1823.5-3020 & 534.1  & 0.027  & 1.41 $\pm$ 1.09  & 1.39 $\pm$ 0.18 \\
J1858-2216 & 4FGL J1858.3-2216 & 586.8  & 0.005  & 1.13 $\pm$ 0.58  & 0.89 $\pm$ 0.05 \\
J1959+2048 & 4FGL J1959.5+2048 & 615.5  & 0.009  & 1.58 $\pm$ 0.54  & 1.97 $\pm$ 0.16 \\
J0418+6635 & 4FGL J0418.9+6636 & 621.9  & 0.014  & 1.10 $\pm$ 0.09  & 0.89 $\pm$ 0.14 \\
J0533+6759 & 4FGL J0533.8+6800 & 635.5  & 0.012  & 0.88 $\pm$ 0.06  & 0.85 $\pm$ 0.10 \\
J2234+0944 & 4FGL J2234.7+0943 & 643.9  & 0.012  & 1.07 $\pm$ 0.08  & 0.82 $\pm$ 0.11 \\
J1824-2452A & 4FGL J1824.6-2452 & 718.7  & 0.028  & 2.10 $\pm$ 1.07  & 2.85 $\pm$ 0.27 \\
J0751+1807 & 4FGL J0751.2+1808 & 819.4  & 0.016  & 1.04 $\pm$ 0.08  & 0.63 $\pm$ 0.10 \\
J1946-5403 & 4FGL J1946.5-5402 & 826.3  & 0.011  & 1.04 $\pm$ 0.06  & 1.21 $\pm$ 0.13 \\
J1302-3258 & 4FGL J1302.4-3258 & 848.1  & 0.015  & 1.12 $\pm$ 0.77  & 0.73 $\pm$ 0.20 \\
J1658-5324 & 4FGL J1658.6-5323 & 985.9  & 0.007  & 2.04 $\pm$ 0.97  & 2.84 $\pm$ 0.50 \\
J0102+4839 & 4FGL J0102.8+4839 & 1036.5  & 0.003  & 1.43 $\pm$ 0.84  & 1.49 $\pm$ 0.15 \\
J1816+4510 & 4FGL J1816.5+4510 & 1043.4  & 0.007  & 1.00 $\pm$ 0.06  & 0.99 $\pm$ 0.09 \\
J1124-3653 & 4FGL J1124.0-3653 & 1060.3  & 0.015  & 1.37 $\pm$ 0.85  & 1.19 $\pm$ 0.05 \\
J1312+0051 & 4FGL J1312.7+0050 & 1204.1  & 0.008  & 1.38 $\pm$ 0.68  & 1.38 $\pm$ 0.10 \\
J2039-5617 & 4FGL J2039.5-5617 & 1205.7  & 0.007  & 1.50 $\pm$ 0.74  & 1.90 $\pm$ 0.09 \\
J2215+5135 & 4FGL J2215.6+5135 & 1211.9  & 0.008  & 1.83 $\pm$ 0.09  & 2.00 $\pm$ 0.15 \\
J1035-6720 & 4FGL J1035.4-6720 & 1485.3  & 0.003  & 1.94 $\pm$ 0.58  & 1.81 $\pm$ 0.09 \\
J0307+7443 & 4FGL J0307.8+7443 & 1560.8  & 0.004  & 1.59 $\pm$ 0.80  & 1.32 $\pm$ 0.25 \\
J0340+4130 & 4FGL J0340.3+4130 & 1577.4  & 0.001  & 1.90 $\pm$ 1.31  & 1.06 $\pm$ 0.18 \\
J1810+1744 & 4FGL J1810.5+1744 & 1610.4  & 0.012  & 2.29 $\pm$ 0.34  & 3.68 $\pm$ 0.28 \\
J1227-4853 & 4FGL J1228.0-4853 & 1625.6  & 0.020  & 2.29 $\pm$ 0.24  & 3.61 $\pm$ 0.34 \\
J1744-7619 & 4FGL J1744.0-7618 & 1784.1  & 0.005  & 1.99 $\pm$ 0.08  & 1.74 $\pm$ 0.12 \\
J0101-6422 & 4FGL J0101.1-6422 & 1822.7  & 0.012  & 1.33 $\pm$ 0.74  & 1.53 $\pm$ 0.13 \\
J1625-0021 & 4FGL J1625.1-0020 & 1878.4  & 0.011  & 2.14 $\pm$ 0.11  & 1.64 $\pm$ 0.15 \\
J0034-0534 & 4FGL J0034.3-0534 & 1961.0  & 0.006  & 2.02 $\pm$ 1.16  & 2.67 $\pm$ 0.39 \\
J1614-2230 & 4FGL J1614.5-2230 & 2177.6  & 0.010  & 2.59 $\pm$ 1.19  & 1.70 $\pm$ 0.08 \\
J1902-5105 & 4FGL J1902.0-5105 & 2290.1  & 0.010  & 2.45 $\pm$ 0.64  & 3.86 $\pm$ 0.17 \\
J1744-1134 & 4FGL J1744.4-1135 & 2328.5  & 0.009  & 3.88 $\pm$ 1.01  & 4.35 $\pm$ 0.20 \\
J1653-0158 & 4FGL J1653.6-0158 & 2422.3  & 0.011  & 3.35 $\pm$ 0.10  & 4.84 $\pm$ 0.22 \\
J2043+1711 & 4FGL J2043.3+1711 & 2478.1  & 0.004  & 2.66 $\pm$ 0.66  & 2.66 $\pm$ 0.21 \\
J0613-0200 & 4FGL J0613.7-0201 & 2678.6  & 0.014  & 3.82 $\pm$ 0.12  & 4.85 $\pm$ 0.22 \\
J0437-4715 & 4FGL J0437.2-4715 & 3019.1  & 0.006  & 1.73 $\pm$ 0.06  & 2.74 $\pm$ 0.13 \\
J2339-0533 & 4FGL J2339.6-0533 & 3058.5  & 0.001  & 2.73 $\pm$ 0.12  & 2.55 $\pm$ 0.15 \\
J2241-5236 & 4FGL J2241.7-5236 & 3063.8  & 0.005  & 2.59 $\pm$ 0.91  & 2.05 $\pm$ 0.27 \\
J1514-4946 & 4FGL J1514.3-4946 & 3105.5  & 0.003  & 4.07 $\pm$ 1.14  & 3.06 $\pm$ 0.13 \\
J2017+0603 & 4FGL J2017.4+0602 & 3174.7  & 0.012  & 3.56 $\pm$ 0.15  & 1.96 $\pm$ 0.13 \\
J2214+3000 & 4FGL J2214.6+3000 & 4809.6  & 0.009  & 3.20 $\pm$ 0.09  & 2.63 $\pm$ 0.11 \\
J2302+4442 & 4FGL J2302.7+4443 & 4991.6  & 0.010  & 3.66 $\pm$ 0.10  & 2.79 $\pm$ 0.11 \\
J2124-3358 & 4FGL J2124.7-3358 & 5935.9  & 0.010  & 3.94 $\pm$ 1.50  & 2.93 $\pm$ 0.43 \\
J1536-4948 & 4FGL J1536.4-4948 & 6600.5  & 0.004  & 7.96 $\pm$ 0.59  & 7.24 $\pm$ 0.24 \\
J0218+4232 & 4FGL J0218.1+4232 & 6771.2  & 0.007  & 4.97 $\pm$ 0.10  & 8.32 $\pm$ 0.22 \\
J1311-3430 & 4FGL J1311.7-3430 & 9011.3  & 0.007  & 6.23 $\pm$ 0.38  & 8.49 $\pm$ 0.26 \\
J0030+0451 & 4FGL J0030.4+0451 & 10547.1  & 0.001  & 5.89 $\pm$ 1.53  & 6.21 $\pm$ 0.33 \\
J1231-1411 & 4FGL J1231.1-1412 & 18868.5  & 0.002  & 10.20 $\pm$ 0.18  & 7.61 $\pm$ 0.16 \\
J0614-3329 & 4FGL J0614.1-3329 & 26170.5  & 0.003  & 11.34 $\pm$ 0.20  & 7.87 $\pm$ 0.14 \\
		\hline

    	\caption{
    	Analysis results for all significantly detected MSPs, in the energy range 100 MeV$-$100 GeV , with 4FGL source id, detection significance (TS),  offset from catalogue co-ordinates and fluxes.}
        \label{tab:MSP_PL_FLUXES}
\end{longtable}

\section{Detected MSP Spectral Parameters and exemplar spectra}
\label{sec:SpectralParamsSEDs}
\begin{longtable}{lc c c c c  }
  	\hline
   MSP Name	&	Prefactor			&	            Index1			&	Scale	&	Expfactor 					\\  
         	&	10\textsuperscript{-13}			&	          			&	&	10\textsuperscript{-3}					          	\\       
    \hline
    \endfirsthead
    \multicolumn{3}{@{}l}{\ldots continued}\\\hline

   MSP Name	&	Prefactor			&	            Index1			&	Scale	&	Expfactor 					\\  
         	&	10\textsuperscript{-13}			&	           			&		&	10\textsuperscript{-3}\\
    \hline
    \endhead
    \hline
    \multicolumn{3}{r@{}}{continued \ldots}\\
    \endfoot
    \hline
    \endlastfoot
J1921+1929	&	7.48	$\pm$	8.30	&	-1.16	$\pm$	0.31	&	3061	&	11.46	$\pm$	0.00	\\
J1137+7528	&	3.46	$\pm$	14.25	&	0.00	$\pm$	0.01	&	3077	&	13.03	$\pm$	4.65	\\
J1125-6014	&	22.07	$\pm$	69.17	&	-0.01	$\pm$	1.08	&	2576	&	15.66	$\pm$	0.39	\\
J1833-3840	&	17.35	$\pm$	42.02	&	-0.57	$\pm$	1.73	&	1538	&	14.85	$\pm$	0.00	\\
J0931-1902	&	25.90	$\pm$	74.11	&	0.00	$\pm$	0.01	&	1926	&	19.91	$\pm$	0.05	\\
J0653+4706	&	66.48	$\pm$	51.65	&	0.00	$\pm$	0.03	&	1321	&	26.28	$\pm$	5.31	\\
J0248+4230	&	25.53	$\pm$	61.13	&	-0.35	$\pm$	1.24	&	1961	&	19.33	$\pm$	0.04	\\
J1855-1436	&	5.21	$\pm$	11.29	&	-1.24	$\pm$	1.26	&	2051	&	5.68	$\pm$	0.00	\\
J0952-0607	&	29.92	$\pm$	66.18	&	-0.20	$\pm$	1.78	&	1583	&	18.71	$\pm$	0.05	\\
J1730-2304	&	156.25	$\pm$	207.42	&	-0.69	$\pm$	0.87	&	1251	&	22.97	$\pm$	0.00	\\
J1908+2105	&	6.36	$\pm$	3.59	&	-1.34	$\pm$	0.23	&	2223	&	7.13	$\pm$	0.15	\\
J2051-0827	&	21.52	$\pm$	48.11	&	0.00	$\pm$	0.01	&	2105	&	16.05	$\pm$	0.16	\\
J1640+2224	&	9.66	$\pm$	18.85	&	-1.05	$\pm$	1.23	&	1763	&	11.67	$\pm$	0.07	\\
J1544+4937	&	4.43	$\pm$	0.99	&	-1.47	$\pm$	0.20	&	1193	&	5.00	$\pm$	0.01	\\
J0621+2514	&	33.52	$\pm$	52.22	&	-0.27	$\pm$	1.13	&	2653	&	15.36	$\pm$	0.02	\\
J2006+0148	&	4.76	$\pm$	1.75	&	-0.84	$\pm$	0.26	&	2658	&	7.44	$\pm$	0.00	\\
J2039-3616	&	17.97	$\pm$	29.87	&	-1.20	$\pm$	1.03	&	1300	&	11.88	$\pm$	0.77	\\
J1641+8049	&	8.35	$\pm$	4.13	&	-1.05	$\pm$	0.33	&	1663	&	11.65	$\pm$	0.02	\\
J2047+1053	&	5.50	$\pm$	5.97	&	-1.17	$\pm$	0.72	&	1964	&	6.41	$\pm$	0.01	\\
J1732-5049	&	14.56	$\pm$	24.84	&	-1.23	$\pm$	1.01	&	1684	&	9.65	$\pm$	0.01	\\
J1741+1351	&	12.33	$\pm$	3.46	&	-0.97	$\pm$	0.23	&	1372	&	8.62	$\pm$	1.79	\\
J1125-5825	&	11.91	$\pm$	20.67	&	-0.71	$\pm$	0.98	&	3232	&	10.33	$\pm$	0.09	\\
J1805+0615	&	27.56	$\pm$	43.89	&	-0.54	$\pm$	1.13	&	1750	&	13.16	$\pm$	0.02	\\
J1939+2134	&	53.65	$\pm$	14.00	&	-1.38	$\pm$	0.12	&	1892	&	12.35	$\pm$	0.01	\\
J1552+5437	&	7.19	$\pm$	5.90	&	-1.14	$\pm$	0.64	&	1318	&	7.66	$\pm$	0.04	\\
J1824+1014	&	13.25	$\pm$	15.49	&	-1.02	$\pm$	0.94	&	1619	&	8.16	$\pm$	0.01	\\
J0740+6620	&	10.19	$\pm$	39.13	&	-1.11	$\pm$	2.75	&	1340	&	9.45	$\pm$	0.05	\\
J1036-8317	&	19.50	$\pm$	2.14	&	-0.63	$\pm$	0.14	&	1787	&	12.37	$\pm$	0.00	\\
J1024-0719	&	22.60	$\pm$	32.71	&	-0.99	$\pm$	0.99	&	1305	&	12.12	$\pm$	0.04	\\
J2115+5448	&	11.10	$\pm$	5.30	&	-0.80	$\pm$	0.23	&	3387	&	8.84	$\pm$	0.10	\\
J1012-4235	&	8.50	$\pm$	0.95	&	-1.27	$\pm$	0.12	&	1849	&	7.45	$\pm$	0.01	\\
J1921+0137	&	15.57	$\pm$	4.21	&	-1.34	$\pm$	0.17	&	1871	&	7.16	$\pm$	0.01	\\
J1207-5050	&	38.90	$\pm$	31.00	&	-0.55	$\pm$	0.45	&	1950	&	15.72	$\pm$	4.66	\\
J0955-6150	&	25.58	$\pm$	23.25	&	-1.23	$\pm$	0.63	&	1256	&	9.56	$\pm$	0.01	\\
J0251+2606	&	67.61	$\pm$	70.78	&	-0.46	$\pm$	0.87	&	1361	&	18.64	$\pm$	0.00	\\
J1446-4701	&	25.87	$\pm$	31.74	&	-0.95	$\pm$	0.91	&	1450	&	11.01	$\pm$	0.26	\\
J1827-0849	&	15.66	$\pm$	19.05	&	-1.55	$\pm$	0.69	&	2349	&	5.00	$\pm$	0.02	\\
J1048+2339	&	15.05	$\pm$	2.95	&	-1.85	$\pm$	0.14	&	1013	&	5.77	$\pm$	1.50	\\
J2042+0246	&	235.13	$\pm$	219.24	&	-0.39	$\pm$	0.81	&	1080	&	27.00	$\pm$	0.01	\\
J1543-5149	&	40.13	$\pm$	45.76	&	-1.48	$\pm$	0.65	&	1582	&	9.57	$\pm$	0.00	\\
J2017-1614	&	12.20	$\pm$	12.01	&	-1.35	$\pm$	0.78	&	1608	&	7.28	$\pm$	0.01	\\
J0318+0253	&	42.08	$\pm$	56.41	&	-0.74	$\pm$	0.82	&	1645	&	15.32	$\pm$	0.20	\\
J1843-1113	&	90.51	$\pm$	97.67	&	-1.59	$\pm$	0.68	&	1017	&	11.42	$\pm$	0.02	\\
J1649-3012	&	48.30	$\pm$	48.51	&	-1.03	$\pm$	0.58	&	1512	&	12.89	$\pm$	0.05	\\
J1622-0315	&	17.45	$\pm$	17.08	&	-1.50	$\pm$	0.76	&	1433	&	6.69	$\pm$	0.02	\\
J0605+3757	&	97.79	$\pm$	9.51	&	0.00	$\pm$	0.00	&	1904	&	19.08	$\pm$	0.01	\\
J2129-0429	&	34.71	$\pm$	38.50	&	-0.83	$\pm$	0.86	&	1186	&	12.28	$\pm$	0.06	\\
J1713+0747	&	32.93	$\pm$	8.76	&	-0.84	$\pm$	0.19	&	1603	&	11.78	$\pm$	1.67	\\
J1513-2550	&	38.30	$\pm$	41.98	&	-0.97	$\pm$	0.74	&	1375	&	12.25	$\pm$	0.02	\\
J1221-0633	&	32.53	$\pm$	35.06	&	-1.19	$\pm$	0.79	&	1123	&	10.44	$\pm$	0.30	\\
J1903-7051	&	16.52	$\pm$	6.18	&	-0.90	$\pm$	0.24	&	2071	&	10.69	$\pm$	1.99	\\
J1628-3205	&	77.18	$\pm$	57.57	&	-1.09	$\pm$	0.40	&	1400	&	15.21	$\pm$	0.04	\\
J1142+0119	&	12.33	$\pm$	2.42	&	-1.27	$\pm$	0.17	&	1436	&	6.32	$\pm$	1.22	\\
J1901-0125	&	43.79	$\pm$	43.07	&	-1.48	$\pm$	0.61	&	1453	&	8.08	$\pm$	0.02	\\
J2310-0555	&	20.35	$\pm$	7.16	&	-1.50	$\pm$	0.22	&	1207	&	7.23	$\pm$	2.55	\\
J1747-4036	&	71.71	$\pm$	68.54	&	-1.10	$\pm$	0.57	&	1471	&	14.04	$\pm$	0.00	\\
J0610-2100	&	49.28	$\pm$	46.63	&	-0.91	$\pm$	1.04	&	1182	&	13.79	$\pm$	1.49	\\
J1600-3053	&	25.06	$\pm$	22.43	&	-0.54	$\pm$	0.62	&	2399	&	10.96	$\pm$	0.01	\\
J0023+0923	&	62.43	$\pm$	12.22	&	-1.05	$\pm$	0.15	&	973	&	12.70	$\pm$	1.66	\\
J1745+1017	&	56.02	$\pm$	55.83	&	-0.65	$\pm$	0.71	&	1532	&	13.97	$\pm$	0.01	\\
J1630+3734	&	68.71	$\pm$	70.94	&	-0.30	$\pm$	0.82	&	1465	&	18.06	$\pm$	0.02	\\
J1301+0833	&	56.21	$\pm$	19.61	&	-0.96	$\pm$	0.25	&	1077	&	12.94	$\pm$	0.02	\\
J2256-1024	&	44.90	$\pm$	39.07	&	-1.20	$\pm$	0.76	&	981	&	10.32	$\pm$	0.00	\\
J1823-3021A	&	20.72	$\pm$	20.29	&	-1.30	$\pm$	0.53	&	1940	&	7.23	$\pm$	0.06	\\
J1858-2216	&	120.63	$\pm$	81.44	&	-0.36	$\pm$	0.50	&	1794	&	17.55	$\pm$	0.01	\\
J1959+2048	&	103.95	$\pm$	48.94	&	-1.05	$\pm$	0.28	&	1308	&	14.14	$\pm$	0.00	\\
J0533+6759	&	16.26	$\pm$	3.02	&	-1.32	$\pm$	0.12	&	1536	&	6.78	$\pm$	1.12	\\
J2234+0944	&	31.34	$\pm$	6.17	&	-0.92	$\pm$	0.14	&	1647	&	9.70	$\pm$	1.19	\\
J1824-2452A	&	75.64	$\pm$	56.68	&	-1.34	$\pm$	0.44	&	1399	&	10.75	$\pm$	0.01	\\
J0751+1807	&	57.62	$\pm$	18.00	&	-0.37	$\pm$	0.21	&	2271	&	14.00	$\pm$	1.61	\\
J1946-5403	&	68.30	$\pm$	20.78	&	-0.95	$\pm$	0.17	&	1472	&	14.72	$\pm$	2.07	\\
J1302-3258	&	53.93	$\pm$	41.62	&	-0.55	$\pm$	0.88	&	1667	&	12.51	$\pm$	0.64	\\
J1658-5324	&	275.92	$\pm$	182.74	&	-0.91	$\pm$	0.38	&	1136	&	19.04	$\pm$	0.02	\\
J0102+4839	&	30.06	$\pm$	19.17	&	-1.39	$\pm$	0.51	&	1362	&	6.65	$\pm$	0.00	\\
J1816+4510	&	42.73	$\pm$	5.21	&	-1.09	$\pm$	0.10	&	1211	&	10.15	$\pm$	0.89	\\
J1124-3653	&	42.93	$\pm$	26.38	&	-1.06	$\pm$	0.50	&	1391	&	9.15	$\pm$	0.06	\\
J1312+0051	&	109.13	$\pm$	68.36	&	-0.77	$\pm$	0.54	&	1353	&	15.08	$\pm$	0.01	\\
J2039-5617	&	43.88	$\pm$	24.27	&	-1.53	$\pm$	0.43	&	1098	&	6.85	$\pm$	0.01	\\
J2215+5135	&	29.76	$\pm$	2.36	&	-1.45	$\pm$	0.06	&	1565	&	6.31	$\pm$	0.46	\\
J1035-6720	&	135.86	$\pm$	53.04	&	-0.71	$\pm$	0.25	&	1615	&	15.01	$\pm$	0.00	\\
J0307+7443	&	182.07	$\pm$	106.38	&	-0.41	$\pm$	0.35	&	1549	&	17.79	$\pm$	0.03	\\
J0340+4130	&	55.67	$\pm$	32.52	&	-0.66	$\pm$	0.43	&	1637	&	9.82	$\pm$	0.00	\\
J1810+1744	&	108.13	$\pm$	20.20	&	-1.61	$\pm$	0.12	&	944	&	9.08	$\pm$	0.01	\\
J1227-4853	&	25.71	$\pm$	3.91	&	-1.82	$\pm$	0.09	&	1597	&	5.29	$\pm$	0.00	\\
J1744-7619	&	181.73	$\pm$	29.36	&	-0.60	$\pm$	0.11	&	1273	&	15.64	$\pm$	1.20	\\
J0101-6422	&	110.28	$\pm$	58.12	&	-1.06	$\pm$	0.46	&	898	&	12.69	$\pm$	0.02	\\
J1625-0021	&	111.50	$\pm$	29.82	&	-0.70	$\pm$	0.17	&	1481	&	12.56	$\pm$	1.78	\\
J0034-0534	&	127.77	$\pm$	63.82	&	-1.48	$\pm$	0.38	&	732	&	8.06	$\pm$	0.01	\\
J1614-2230	&	183.68	$\pm$	92.25	&	-0.37	$\pm$	0.42	&	1771	&	14.89	$\pm$	0.01	\\
J1902-5105	&	94.38	$\pm$	37.34	&	-1.58	$\pm$	0.29	&	1083	&	9.12	$\pm$	0.00	\\
J1744-1134	&	954.11	$\pm$	300.05	&	-0.45	$\pm$	0.27	&	1033	&	22.73	$\pm$	0.00	\\
J1653-0158	&	215.58	$\pm$	6.77	&	-1.47	$\pm$	0.04	&	820	&	9.61	$\pm$	0.00	\\
J2043+1711	&	86.28	$\pm$	21.19	&	-1.23	$\pm$	0.16	&	1212	&	8.35	$\pm$	0.01	\\
J0613-0200	&	176.02	$\pm$	5.59	&	-1.38	$\pm$	0.03	&	1001	&	9.13	$\pm$	0.00	\\
J0437-4715	&	574.20	$\pm$	57.88	&	-0.88	$\pm$	0.08	&	624	&	23.04	$\pm$	1.21	\\
J2339-0533	&	76.33	$\pm$	7.56	&	-1.33	$\pm$	0.07	&	1063	&	6.36	$\pm$	0.73	\\
J2241-5236	&	144.93	$\pm$	60.61	&	-0.62	$\pm$	0.22	&	1880	&	13.93	$\pm$	0.02	\\
J1514-4946	&	87.27	$\pm$	23.63	&	-1.05	$\pm$	0.17	&	1627	&	7.92	$\pm$	0.00	\\
J2017+0603	&	80.12	$\pm$	9.03	&	-0.75	$\pm$	0.08	&	1741	&	8.67	$\pm$	0.65	\\
J2214+3000	&	280.87	$\pm$	7.96	&	-0.60	$\pm$	0.04	&	1092	&	14.71	$\pm$	0.00	\\
J2302+4442	&	164.61	$\pm$	4.38	&	-0.81	$\pm$	0.03	&	1292	&	11.00	$\pm$	0.00	\\
J2124-3358	&	396.58	$\pm$	131.19	&	-0.42	$\pm$	0.24	&	1125	&	15.96	$\pm$	0.00	\\
J1536-4948	&	123.55	$\pm$	8.81	&	-1.39	$\pm$	0.05	&	1445	&	5.38	$\pm$	0.00	\\
J0218+4232	&	226.13	$\pm$	4.97	&	-1.80	$\pm$	0.02	&	802	&	6.54	$\pm$	0.00	\\
J1311-3430	&	263.06	$\pm$	15.13	&	-1.64	$\pm$	0.04	&	812	&	6.04	$\pm$	0.00	\\
J0030+0451	&	577.90	$\pm$	141.06	&	-0.87	$\pm$	0.21	&	930	&	14.43	$\pm$	0.00	\\
J1231-1411	&	692.13	$\pm$	11.07	&	-0.65	$\pm$	0.02	&	1057	&	12.83	$\pm$	0.00	\\
J0614-3329	&	337.66	$\pm$	4.79	&	-1.03	$\pm$	0.02	&	1070	&	7.40	$\pm$	0.00	\\

    		\hline

    	\caption{The spectral parameters for 108 MSPs, in the energy range 100 MeV$-$100 GeV, described by a PLSuperExp2 spectral model, in order of increasing detection significance. The \textit{index2} parameter (not shown) has value 0.67 throughout. }
        \label{tab:MSP_PLSUPEREXP_2_MODELS}
\end{longtable}

\begin{longtable}{lc c c c c  }
  	\hline
        MSP Name	&	Norm			&	            Alpha			&	Beta	& Eb         	   \\	&10\textsuperscript{-13}		&	            			    &   \\       

    \hline
    \endfirsthead
    \multicolumn{3}{@{}l}{\ldots continued}\\
    \hline
            MSP Name	&	Norm			&	            Alpha			&	Beta	& Eb\\          	   	&10\textsuperscript{-13}		&	            			    &   \\       

    \hline
    \endhead
    \hline
    \multicolumn{3}{r@{}}{continued \ldots}\\
    \endfoot
    \hline
    \endlastfoot
J1335-5656	&	6.95	$\pm$	2.49	&	2.06	$\pm$	0.13	&	0.29	$\pm$	0.09	&	1547	\\
J1400-1431	&	14.54	$\pm$	1.53	&	2.13	$\pm$	0.12	&	0.34	$\pm$	0.10	&	987	\\
J0312-0921	&	12.35	$\pm$	1.42	&	1.80	$\pm$	0.19	&	0.94	$\pm$	0.24	&	1179	\\
J2034+3632	&	6.09	$\pm$	3.89	&	2.11	$\pm$	0.62	&	0.97	$\pm$	0.02	&	2678	\\
J0418+6635	&	9.15	$\pm$	0.66	&	2.04	$\pm$	0.08	&	0.29	$\pm$	0.06	&	1525	\\

        		\hline

    	\caption{The spectral parameters for MSPs, in the energy range 100 MeV$-$100 GeV, with a LP spectral model, in order of increasing detection significance. }
        \label{tab:MSP_LP_MODELS}
\end{longtable}

\begin{longtable}{lc c c c c c }
  	\hline
   MSP Name	&	Prefactor			&	            Index1			&	Scale	\\       
   	&	10\textsuperscript{-13}		&	            			    &		    \\       
   
    \hline
    \endfirsthead
    \multicolumn{3}{@{}l}{\ldots continued}\\\hline
    MSP Name	&	Prefactor			&	            Index1			&	Scale	\\       
   	&	10\textsuperscript{-13}		&	            			    &		    \\       
   
    \hline
    \endhead
    \hline
    \multicolumn{3}{r@{}}{continued \ldots}\\
    \endfoot
    \hline
    \endlastfoot

J0737-3039A	&	3.02	$\pm$	0.82	&	-1.98	$\pm$	0.15	&	1000	\\
J1811-2405	&	0.78	$\pm$	0.18	&	-2.22	$\pm$	0.12	&	3126	\\
J2052+1218	&	3.68	$\pm$	0.55	&	-2.27	$\pm$	0.11	&	1165	\\
J1431-4715	&	8.07	$\pm$	1.09	&	-2.38	$\pm$	0.10	&	868	\\
J1555-2908	&	10.26	$\pm$	1.27	&	-2.31	$\pm$	0.09	&	801	\\

        		\hline

    	\caption{The spectral parameters for MSPs, in the energy range 100 MeV$-$100 GeV, with a PL spectral model, in order of increasing detection significance. }
        \label{tab:MSP_PL_MODELS}
\end{longtable}




\begin{figure}
\includegraphics[width=0.49\textwidth]{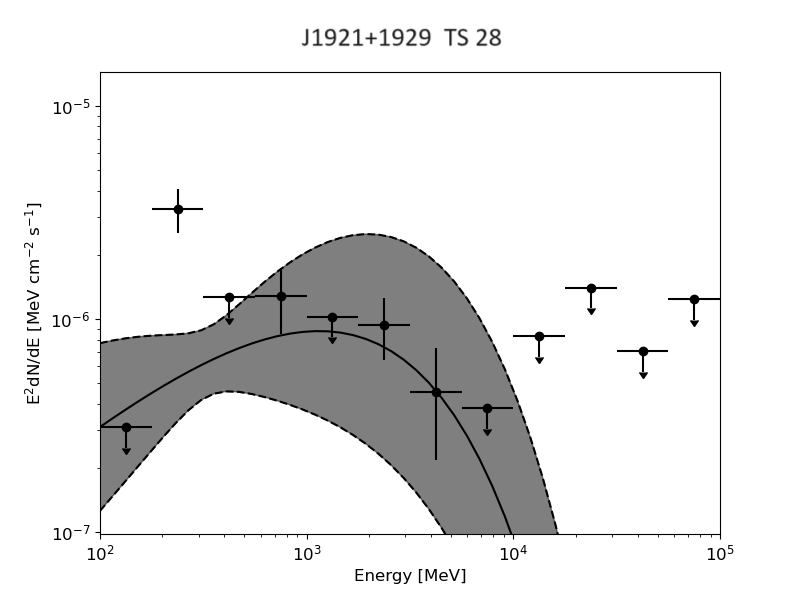}
\includegraphics[width=0.49\textwidth]{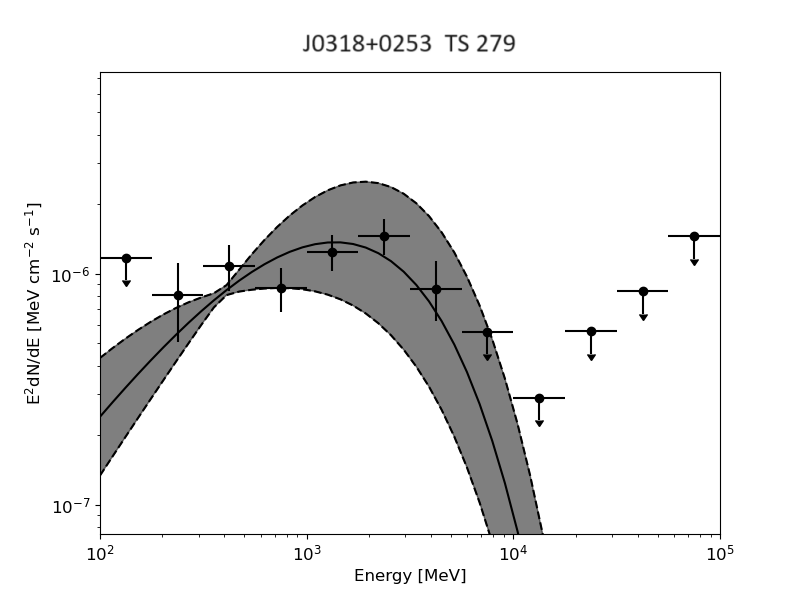}
\includegraphics[width=0.49\textwidth]{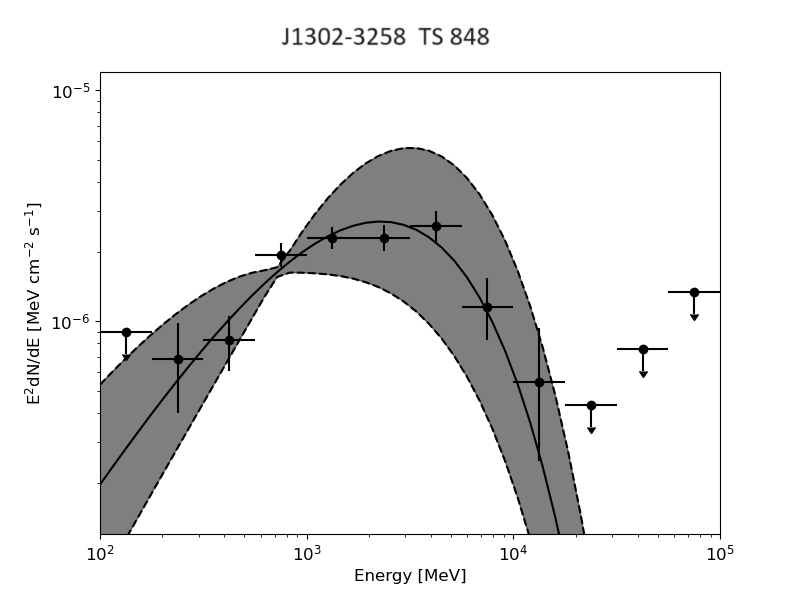}
\includegraphics[width=0.49\textwidth]{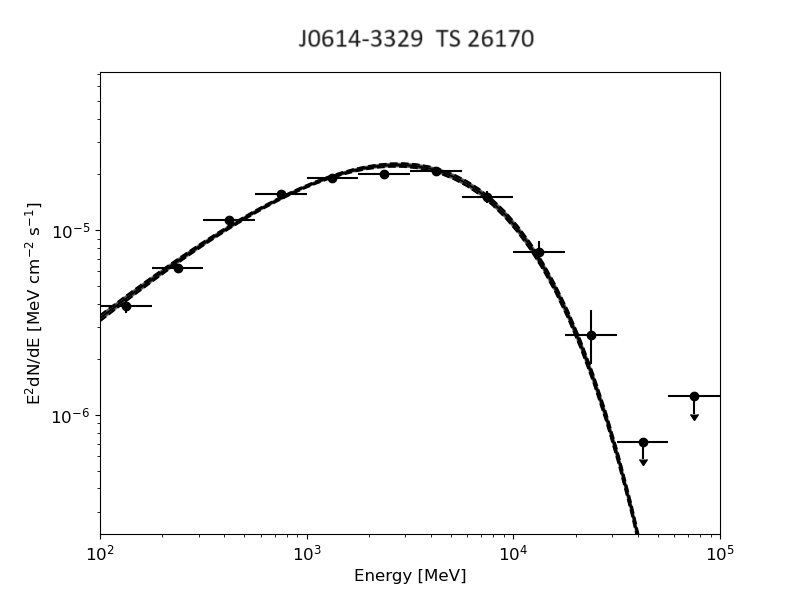}

\caption{ Individual MSP spectra fitted with a PLSuperExp2 spectral model. The examples here range from the lowest to highest detection significance (TS).}
\label{fig:PLSUP_EXP_2_SEDS}   
\end{figure}

\begin{figure}
\includegraphics[width=0.49\textwidth]{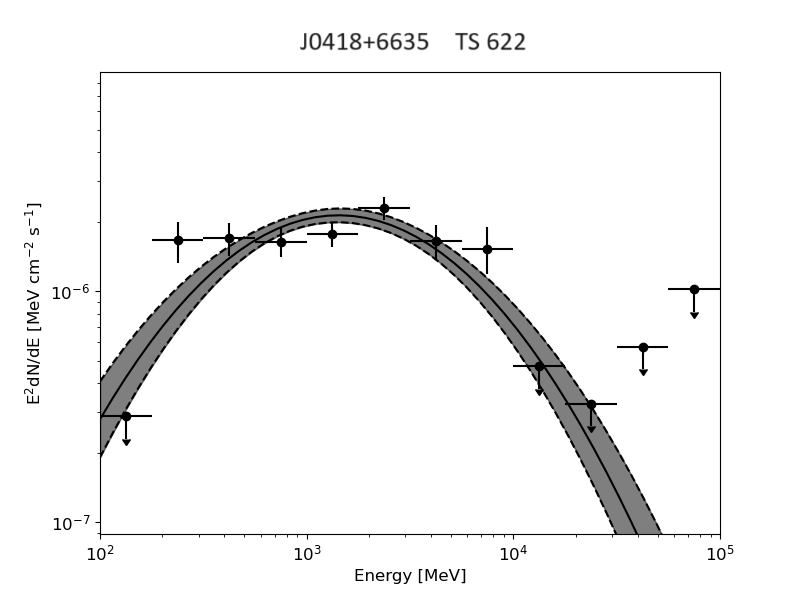}
\includegraphics[width=0.49\textwidth]{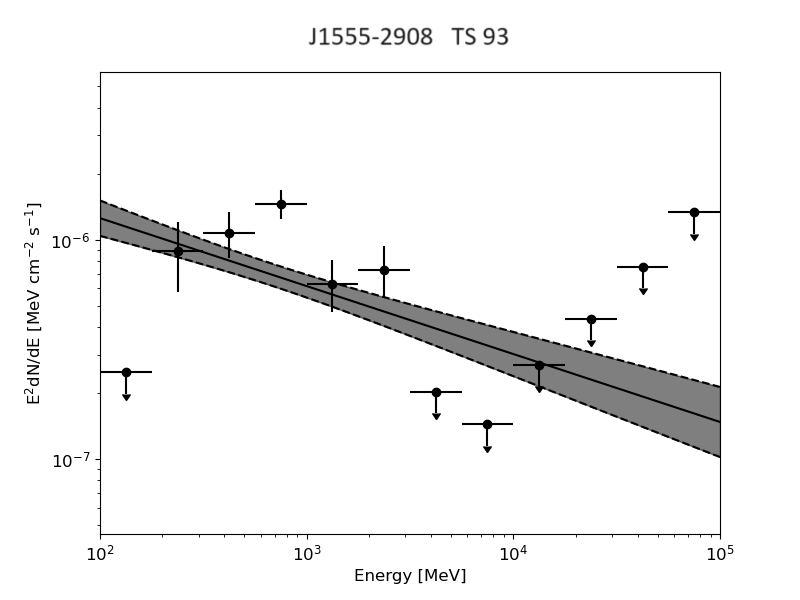}
\caption{ Individual MSP spectra fitted with a log parabola (left) and power law model (right). The examples here are those with the highest detection significance.} 
\label{fig:PL_LP_SED}   
\end{figure}

\section{List of Selected MSPs}

\begin{longtable}{lc c c c c }

    \hline
MSP Name	&	 RA	&	 DEC	&	Period	&$\dot{E}$		\\
            &   (Deg.)  &(Deg.)     &    (ms)   & 10\textsuperscript{34} erg s\textsuperscript{-1}\\
    \hline
    \endfirsthead
    \multicolumn{3}{@{}l}{\ldots continued}\\\hline
MSP Name	&	 RA	&	 DEC	&	Period	&$\dot{E}$		\\
            &   (Deg.)  &(Deg.)     &    (ms)   & 10\textsuperscript{34} erg s\textsuperscript{-1}\\
    
    \hline
    \endhead
    \hline
    \multicolumn{3}{r@{}}{continued \ldots}\\
    \endfoot
    \hline
    \endlastfoot

 J0931-1902  & 142.83 & -19.05 & 4.64 & 0.14 \\
 J1730-2304  & 262.59 & -23.08 & 8.12 & 0.15 \\
 J1455-3330  & 223.95 & -33.51 & 7.99 & 0.19 \\
 J1207-5050  & 181.84 & -50.84 & 4.84 & 0.21 \\
 J2317+1439  & 349.29 & 14.66 & 3.45 & 0.23 \\
 J1640+2224  & 250.07 & 22.4 & 3.16 & 0.35 \\
 J1713+0747  & 258.46 & 7.79 & 4.57 & 0.35 \\
 J0030+0451  & 7.61 & 4.86 & 4.87 & 0.35 \\
 J1327-0755  & 201.99 & -7.93 & 2.68 & 0.37 \\
 J1744-7619  & 266 & -76.32 & 4.69 & 0.37 \\
 J2302+4442  & 345.7 & 44.71 & 5.19 & 0.37 \\
 J1732-5049  & 263.2 & -50.82 & 5.31 & 0.37 \\
 J1946+3417  & 296.6 & 34.29 & 3.17 & 0.39 \\
 J1142+0119  & 175.71 & 1.33 & 5.08 & 0.45 \\
 J1744-1134  & 266.12 & -11.58 & 4.07 & 0.52 \\
 J1745+1017  & 266.39 & 10.3 & 2.65 & 0.53 \\
 J1024-0719  & 156.16 & -7.32 & 5.16 & 0.53 \\
 J1946-5403  & 296.64 & -54.06 & 2.71 & 0.54 \\
 J2051-0827  & 312.78 & -8.46 & 4.51 & 0.55 \\
 J0636+5129  & 99.02 & 51.48 & 2.87 & 0.57 \\
 J0533+6759  & 83.48 & 67.99 & 4.39 & 0.57 \\
 J0737-3039A  & 114.46 & -30.66 & 22.7 & 0.59 \\
 J2042+0246  & 310.5 & 2.8 & 4.53 & 0.6 \\
 J2124-3358  & 321.18 & -33.98 & 4.93 & 0.69 \\
 J0751+1807  & 117.79 & 18.13 & 3.48 & 0.73 \\
 J0340+4130  & 55.1 & 41.51 & 3.3 & 0.75 \\
 J1552+5437  & 238.22 & 54.62 & 2.43 & 0.77 \\
 J2017-1614  & 304.44 & -16.24 & 2.31 & 0.78 \\
 J1137+7528  & 174.26 & 75.47 & 2.51 & 0.79 \\
 J1622-0315  & 245.75 & -3.26 & 3.85 & 0.79 \\
 J1125-6014  & 171.48 & -60.24 & 2.63 & 0.81 \\
 J1600-3053  & 240.22 & -30.9 & 3.6 & 0.81 \\
 J0610-2100  & 92.56 & -21.01 & 3.86 & 0.84 \\
 J1012-4235  & 153.05 & -42.6 & 3.1 & 0.87 \\
 J1855-1436  & 283.98 & -14.6 & 3.59 & 0.93 \\
 J1312+0051  & 198.19 & 0.85 & 4.23 & 0.93 \\
 J0605+3757  & 91.27 & 37.96 & 2.73 & 0.95 \\
 J1400-1431  & 210.15 & -14.53 & 3.08 & 0.97 \\
 J1858-2216  & 284.57 & -22.28 & 2.38 & 1.1 \\
 J0101-6422  & 15.3 & -64.38 & 2.57 & 1.1 \\
 J2310-0555  & 347.53 & -5.93 & 2.61 & 1.1 \\
 J1630+3734  & 247.65 & 37.58 & 3.32 & 1.1 \\
 J2047+1053  & 311.79 & 10.89 & 4.29 & 1.1 \\
 J1653-0158  & 253.41 & -1.98 & 1.97 & 1.2 \\
 J1544+4937  & 236.02 & 49.63 & 2.16 & 1.2 \\
 J1048+2339  & 162.18 & 23.66 & 4.67 & 1.2 \\
 J0437-4715  & 69.32 & -47.25 & 5.76 & 1.2 \\
 J2017+0603  & 304.34 & 6.05 & 2.9 & 1.3 \\
 J1614-2230  & 243.65 & -22.51 & 3.15 & 1.3 \\
 J2043+1711  & 310.84 & 17.19 & 2.38 & 1.5 \\
 J0023+0923  & 5.82 & 9.39 & 3.05 & 1.5 \\
 J0613-0200  & 93.43 & -2.01 & 3.06 & 1.5 \\
 J1514-4946  & 228.58 & -49.77 & 3.59 & 1.6 \\
 J2234+0944  & 338.7 & 9.74 & 3.63 & 1.6 \\
 J1124-3653  & 171 & -36.89 & 2.41 & 1.7 \\
 J1832-0836  & 278.11 & -8.62 & 2.72 & 1.7 \\
 J0102+4839  & 15.71 & 48.66 & 2.96 & 1.7 \\
 J2214+3000  & 333.66 & 30.01 & 3.12 & 1.8 \\
 J1231-1411  & 187.8 & -14.2 & 3.68 & 1.8 \\
 J0740+6620  & 115.19 & 66.34 & 2.89 & 2 \\
 J2339-0533  & 354.91 & -5.55 & 2.88 & 2.2 \\
 J1909-3744  & 287.45 & -37.74 & 2.95 & 2.2 \\
 J0614-3329  & 93.54 & -33.5 & 3.15 & 2.2 \\
 J0307+7443  & 46.98 & 74.72 & 3.16 & 2.2 \\
 J1741+1351  & 265.38 & 13.86 & 3.75 & 2.2 \\
 J2241-5236  & 340.43 & -52.61 & 2.19 & 2.5 \\
 J1811-2405  & 272.83 & -24.09 & 2.66 & 2.8 \\
 J0034-0534  & 8.59 & -5.58 & 1.88 & 2.9 \\
 J1536-4948  & 234.1 & -49.82 & 3.08 & 2.9 \\
 J2039-5617  & 309.9 & -56.29 & 2.65 & 3 \\
 J1036-8317  & 159.17 & -83.3 & 3.41 & 3 \\
 J1658-5324  & 254.66 & -53.4 & 2.44 & 3.2 \\
 J1446-4701  & 221.65 & -47.02 & 2.19 & 3.8 \\
 J1827-0849  & 276.9 & -8.83 & 2.24 & 3.8 \\
 J0248+4230  & 42.13 & 42.51 & 2.6 & 3.8 \\
 J1810+1744  & 272.66 & 17.74 & 1.66 & 4 \\
 J0621+2514  & 95.3 & 25.23 & 2.72 & 4.8 \\
 J1921+0137  & 290.38 & 1.62 & 2.5 & 4.9 \\
 J1311-3430  & 197.94 & -34.51 & 2.56 & 4.9 \\
 J2215+5135  & 333.89 & 51.59 & 2.61 & 5.2 \\
 J1816+4510  & 274.15 & 45.18 & 3.19 & 5.2 \\
 J1843-1113  & 280.92 & -11.23 & 1.85 & 6 \\
 J1901-0125  & 285.39 & -1.42 & 2.79 & 6.5 \\
 J0952-0607  & 148.03 & -6.12 & 1.41 & 6.7 \\
 J1902-5105  & 285.51 & -51.1 & 1.74 & 6.8 \\
 J1431-4715  & 217.94 & -47.26 & 2.01 & 6.8 \\
 J0955-6150  & 148.83 & -61.84 & 2 & 7 \\
 J1543-5149  & 235.93 & -51.83 & 2.06 & 7.3 \\
 J1035-6720  & 158.86 & -67.34 & 2.87 & 7.7 \\
 J1921+1929  & 290.35 & 19.49 & 2.65 & 8.1 \\
 J1125-5825  & 171.43 & -58.42 & 3.1 & 8.1 \\
 J1227-4853  & 186.99 & -48.9 & 1.69 & 9 \\
 J1513-2550  & 228.35 & -25.84 & 2.12 & 9 \\
 J1903-7051  & 285.91 & -70.86 & 3.6 & 9.9 \\
 J1747-4036  & 266.95 & -40.62 & 1.65 & 12 \\
 J1959+2048  & 299.9 & 20.8 & 1.61 & 16 \\
 J2115+5448  & 318.8 & 54.81 & 2.6 & 17 \\
 J0218+4232  & 34.53 & 42.54 & 2.32 & 24 \\
 J1555-2908  & 238.92 & -29.14 & 1.79 & 31 \\
 J1823-3021A  & 275.92 & -30.36 & 5.44 & 83 \\
 J1939+2134  & 294.91 & 21.58 & 1.56 & 110 \\
 J1824-2452A  & 276.13 & -24.87 & 3.05 & 220 \\
 J1301+0833  & 195.41 & 8.57 & 1.84 & $-$ \\
 J1833-3840  & 278.27 & -38.68 & 1.87 & $-$ \\
 J1221-0633  & 185.35 & -6.56 & 1.93 & $-$ \\
 J2052+1218  & 313.2 & 12.33 & 1.99 & $-$ \\
 J1641+8049  & 250.34 & 80.83 & 2.02 & $-$ \\
 J1805+0615  & 271.43 & 6.26 & 2.13 & $-$ \\
 J2006+0148  & 301.62 & 1.82 & 2.16 & $-$ \\
 J2256-1024  & 344.23 & -10.41 & 2.29 & $-$ \\
 J0154+1833  & 28.65 & 18.56 & 2.36 & $-$ \\
 J2205+6012  & 331.39 & 60.22 & 2.42 & $-$ \\
 J0251+2606  & 42.76 & 26.1 & 2.54 & $-$ \\
 J1908+2105  & 287.24 & 21.08 & 2.56 & $-$ \\
 J1625-0021  & 246.29 & -0.36 & 2.83 & $-$ \\
 J0418+6635  & 64.7 & 66.59 & 2.91 & $-$ \\
 J1628-3205  & 247.03 & -32.1 & 3.21 & $-$ \\
 J1335-5656  & 203.77 & -56.93 & 3.24 & $-$ \\
 J2039-3616  & 309.82 & -36.27 & 3.28 & $-$ \\
 J1649-3012  & 252.44 & -30.21 & 3.42 & $-$ \\
 J2034+3632  & 308.75 & 36.54 & 3.65 & $-$ \\
 J0312-0921  & 48.03 & -9.37 & 3.7 & $-$ \\
 J1302-3258  & 195.61 & -32.98 & 3.77 & $-$ \\
 J1824+1014  & 276.06 & 10.25 & 4.07 & $-$ \\
 J0653+4706  & 103.27 & 47.11 & 4.76 & $-$ \\
 J0318+0253  & 49.56 & 2.88 & 5.19 & $-$ \\
 J2129-0429  & 322.44 & -4.49 & 7.61 & $-$ \\

		\hline

    	\caption{Analysis selection of 127 MSPs from the "Public List of LAT-Detected Gamma-Ray Pulsars" ordered by $\dot{E}$. RAJ and DECJ are right ascension and declination in degrees. 25 MSPs have no $\dot{E}$ given in the online catalogue as indicated by a "$-$". 
    	}
        \label{tab:MSP_SELECTION_LIST}
\end{longtable}







\bibliographystyle{mnras}
\bibliography{exportlist} 

\begin{thebibliography}{}
\makeatletter
\relax
\def\mn@urlcharsother{\let\do\@makeother \do\$\do\&\do\#\do\^\do\_\do\%\do\~}
\def\mn@doi{\begingroup\mn@urlcharsother \@ifnextchar [ {\mn@doi@} {\mn@doi@[]}}
\def\mn@doi@[#1]#2{\def\@tempa{#1}\ifx\@tempa\@empty \href {http://dx.doi.org/#2} {doi:#2}\else \href {http://dx.doi.org/#2} {#1}\fi \endgroup}
\def\mn@eprint#1#2{\mn@eprint@#1:#2::\@nil}
\def\mn@eprint@arXiv#1{\href {http://arxiv.org/abs/#1} {{\tt arXiv:#1}}}
\def\mn@eprint@dblp#1{\href {http://dblp.uni-trier.de/rec/bibtex/#1.xml} {dblp:#1}}
\def\mn@eprint@#1:#2:#3:#4\@nil{\def\@tempa {#1}\def\@tempb {#2}\def\@tempc {#3}\ifx \@tempc \@empty \let \@tempc \@tempb \let \@tempb \@tempa \fi \ifx \@tempb \@empty \def\@tempb {arXiv}\fi \@ifundefined {mn@eprint@\@tempb}{\@tempb:\@tempc}{\expandafter \expandafter \csname mn@eprint@\@tempb\endcsname \expandafter{\@tempc}}}

\bibitem[\protect\citeauthoryear{Abazajian}{Abazajian}{2011}]{RN78}
Abazajian K.~N.,  2011, \mn@doi [Journal of Cosmology and Astroparticle Physics] {10.1088/1475-7516/2011/03/010}

\bibitem[\protect\citeauthoryear{Abdo et~al.,}{Abdo et~al.}{2009a}]{RN196}
Abdo A.~A.,  et~al., 2009a, \mn@doi [Science] {10.1126/science.1177023}, 325, 845

\bibitem[\protect\citeauthoryear{Abdo et~al.,}{Abdo et~al.}{2009b}]{RN82}
Abdo A.~A.,  et~al., 2009b, \mn@doi [Science] {10.1126/science.1176113}, 325, 848

\bibitem[\protect\citeauthoryear{Abdo et~al.,}{Abdo et~al.}{2013}]{RN244}
Abdo A.~A.,  et~al., 2013, \mn@doi [Astrophysical Journal Supplement Series] {10.1088/0067-0049/208/2/17}, 208, 59

\bibitem[\protect\citeauthoryear{Ajello et~al.,}{Ajello et~al.}{2017}]{RN352}
Ajello M.,  et~al., 2017, \mn@doi [Astrophysical Journal Supplement Series] {10.3847/1538-4365/aa8221}, 232

\bibitem[\protect\citeauthoryear{{Akaike}}{{Akaike}}{1974}]{AIC}
{Akaike} H.,  1974, IEEE Transactions on Automatic Control, \href {https://ui.adsabs.harvard.edu/abs/1974ITAC...19..716A} {19, 716}

\bibitem[\protect\citeauthoryear{Bartels \& Edwards}{Bartels \& Edwards}{2019}]{RN739}
Bartels R.,  Edwards T.,  2019, \mn@doi [Physical Review D] {10.1103/PhysRevD.100.068301}, 100

\bibitem[\protect\citeauthoryear{Benli, Petri  \& Mitra}{Benli et~al.}{2021}]{RN635}
Benli O.,  Petri J.,   Mitra D.,  2021, \mn@doi [Astronomy & Astrophysics] {10.1051/0004-6361/202039853}, 647

\bibitem[\protect\citeauthoryear{Brown, Lacroix, Lloyd, Boehm  \& Chadwick}{Brown et~al.}{2018}]{RN430}
Brown A.~M.,  Lacroix T.,  Lloyd S.,  Boehm C.,   Chadwick P.,  2018, \mn@doi [Physical Review D] {10.1103/PhysRevD.98.041301}, 98

\bibitem[\protect\citeauthoryear{Brown, Lacroix, Lloyd, Boehm  \& Chadwick}{Brown et~al.}{2019}]{RN740}
Brown A.~M.,  Lacroix T.,  Lloyd S.,  Boehm C.,   Chadwick P.,  2019, \mn@doi [Physical Review D] {10.1103/PhysRevD.100.068302}, 100

\bibitem[\protect\citeauthoryear{Burnham \& Anderson}{Burnham \& Anderson}{2004}]{RN741}
Burnham K.~P.,  Anderson D.~R.,  2004, \mn@doi [Sociological Methods & Research] {10.1177/0049124104268644}, 33, 261

\bibitem[\protect\citeauthoryear{Cerutti, Philippov  \& Spitkovsky}{Cerutti et~al.}{2016}]{RN522}
Cerutti B.,  Philippov A.~A.,   Spitkovsky A.,  2016, \mn@doi [Monthly Notices of the Royal Astronomical Society] {10.1093/mnras/stw124}, 457, 2401

\bibitem[\protect\citeauthoryear{{Deutsch}}{{Deutsch}}{1955}]{Deutsch}
{Deutsch} A.~J.,  1955, Annales d'Astrophysique, \href {https://ui.adsabs.harvard.edu/abs/1955AnAp...18....1D} {18, 1}

\bibitem[\protect\citeauthoryear{FSSC}{FSSC}{2010}]{RN181}
FSSC 2010, \url {https://fermi.gsfc.nasa.gov/ssc/data/analysis/scitools/source_models.html}

\bibitem[\protect\citeauthoryear{Giraud \& Petri}{Giraud \& Petri}{2021}]{RN716}
Giraud Q.,  Petri J.,  2021, \mn@doi [Astronomy & Astrophysics] {10.1051/0004-6361/202040020}, 654

\bibitem[\protect\citeauthoryear{Guillemot \& Tauris}{Guillemot \& Tauris}{2014}]{RN458}
Guillemot L.,  Tauris T.~M.,  2014, \mn@doi [Monthly Notices of the Royal Astronomical Society] {10.1093/mnras/stu082}, 439, 2033

\bibitem[\protect\citeauthoryear{Guillemot et~al.,}{Guillemot et~al.}{2016}]{RN449}
Guillemot L.,  et~al., 2016, \mn@doi [Astronomy & Astrophysics] {10.1051/0004-6361/201527847}, 587

\bibitem[\protect\citeauthoryear{Johnson et~al.,}{Johnson et~al.}{2014}]{RN28}
Johnson T.~J.,  et~al., 2014, \mn@doi [Astrophysical Journal Supplement Series] {10.1088/0067-0049/213/1/6}, 213

\bibitem[\protect\citeauthoryear{McCann}{McCann}{2015}]{RN197}
McCann A.,  2015, \mn@doi [Astrophysical Journal] {10.1088/0004-637x/804/2/86}, 804, 10

\bibitem[\protect\citeauthoryear{Muslimov \& Harding}{Muslimov \& Harding}{2004}]{Muslimov_2004}
Muslimov A.~G.,  Harding A.~K.,  2004, \mn@doi [The Astrophysical Journal] {10.1086/425227}, 617, 471

\bibitem[\protect\citeauthoryear{Petri}{Petri}{2019}]{RN347}
Petri J.,  2019, \mn@doi [Monthly Notices of the Royal Astronomical Society] {10.1093/mnras/stz360}, 484, 5669

\bibitem[\protect\citeauthoryear{Ploeg, Gordon, Crocker  \& Macias}{Ploeg et~al.}{2020}]{Ploeg_2020}
Ploeg H.,  Gordon C.,  Crocker R.,   Macias O.,  2020, \mn@doi [Journal of Cosmology and Astroparticle Physics] {10.1088/1475-7516/2020/12/035}, 2020, 035

\bibitem[\protect\citeauthoryear{Smith, Guillemot, Kerr, Ng  \& Barr}{Smith et~al.}{2017}]{smith2017gammaray}
Smith D.~A.,  Guillemot L.,  Kerr M.,  Ng C.,   Barr E.,  2017, Gamma-ray pulsars with Fermi (\mn@eprint {arXiv} {1706.03592}), \url {https://arxiv.org/abs/1706.03592}

\bibitem[\protect\citeauthoryear{Smith et~al.,}{Smith et~al.}{2019}]{RN459}
Smith D.~A.,  et~al., 2019, \mn@doi [Astrophysical Journal] {10.3847/1538-4357/aaf57d}, 871

\bibitem[\protect\citeauthoryear{Smith et~al.,}{Smith et~al.}{2023}]{RN823}
Smith D.~A.,  et~al., 2023, \mn@doi [Astrophysical Journal] {10.3847/1538-4357/acee67}, 958

\bibitem[\protect\citeauthoryear{Torres, Vigano, Zelati  \& Li}{Torres et~al.}{2019}]{RN452}
Torres D.~F.,  Vigano D.,  Zelati F.~C.,   Li J.,  2019, \mn@doi [Monthly Notices of the Royal Astronomical Society] {10.1093/mnras/stz2403}, 489, 5494

\bibitem[\protect\citeauthoryear{Vigano, Torres, Hirotani  \& Pessah}{Vigano et~al.}{2015a}]{RN376}
Vigano D.,  Torres D.~F.,  Hirotani K.,   Pessah M.~E.,  2015a, \mn@doi [Monthly Notices of the Royal Astronomical Society] {10.1093/mnras/stu2564}, 447, 2631

\bibitem[\protect\citeauthoryear{Vigano, Torres, Hirotani  \& Pessah}{Vigano et~al.}{2015b}]{RN451}
Vigano D.,  Torres D.~F.,  Hirotani K.,   Pessah M.~E.,  2015b, \mn@doi [Monthly Notices of the Royal Astronomical Society] {10.1093/mnras/stu2565}, 447, 2649

\bibitem[\protect\citeauthoryear{{Wood}, {Caputo}, {Charles}, {Di Mauro}, {Magill}  \& {Jeremy Perkins for the Fermi-LAT Collaboration}}{{Wood} et~al.}{2017}]{2017arXiv170709551W}
{Wood} M.,  {Caputo} R.,  {Charles} E.,  {Di Mauro} M.,  {Magill} J.,   {Jeremy Perkins for the Fermi-LAT Collaboration} 2017, preprint, \href {http://adsabs.harvard.edu/abs/2017arXiv170709551W} {} (\mn@eprint {arXiv} {1707.09551})

\bibitem[\protect\citeauthoryear{Wu, Wang, Xing  \& Zhang}{Wu et~al.}{2022}]{Wu_2022}
Wu W.,  Wang Z.,  Xing Y.,   Zhang P.,  2022, \mn@doi [The Astrophysical Journal] {10.3847/1538-4357/ac4f48}, 927, 117

\bibitem[\protect\citeauthoryear{Xing \& Wang}{Xing \& Wang}{2016}]{RN183}
Xing Y.,  Wang Z.~X.,  2016, \mn@doi [Astrophysical Journal] {10.3847/0004-637x/831/2/143}, 831, 8

\bibitem[\protect\citeauthoryear{Zhang \& Cheng}{Zhang \& Cheng}{2003}]{RN731}
Zhang L.,  Cheng K.~S.,  2003, \mn@doi [Astronomy & Astrophysics] {10.1051/0004-6361:20021570}, 398, 639

\makeatother
\end{thebibliography}

\bsp	
\label{lastpage}
\end{document}